\documentclass[usenatbib]{mn2e}

\usepackage{graphicx}
\usepackage{subfigure}
\usepackage{amssymb}

\newcommand\jbt{{{\bf J}\cdot{\bf B},tot}}

\newcommand\abt{{{\bf A} \cdot{\bf B},tot}}

\newcommand\bbt{{B^2,tot}}
\newcommand\bbs{{B^2,s}}
\newcommand\vet{{v^2,tot}}
\newcommand\ves{{v^2,s}}
\newcommand\jbs{{{\bf J}\cdot{\bf B},s}}

\newcommand\abs{{{\bf A}\cdot{\bf B},s}}

\newcommand\bfe{{\bf e}}

\newcommand\bfb{{\bf b}}
\newcommand\bfv{{\bf v}}

\newcommand\beq{\begin{equation}}
\newcommand\eeq{\end{equation}}

\title[Simulations of a Magnetic Fluctuation Driven Large Scale Dynamo  and Comparison with
a Two-scale Model]{Simulations of a Magnetic Fluctuation Driven Large Scale Dynamo  and Comparison with
a Two-scale Model}
\author[Kiwan Park and E.G. Blackman]{Kiwan Park$^{1}$\thanks{E-mail: pkiwan@pas.rochester.edu}
and E.G. Blackman$^{1}$\thanks{E-mail: blackman@pas.rochester.edu }\\
$^{1}$Department of Physics and Astronomy, University of Rochester, Rochester, NY 14627, USA}
\begin{document}
\date{}
\maketitle
\label{firstpage}

\begin{abstract}
Models  of large scale (magnetohydrodynamic) dynamos  (LSD) which couple large scale  field growth to total magnetic helicity evolution best predict the saturation of LSDs seen in simulations.
For the simplest so called ``$\alpha^2$'' LSDs in periodic boxes,
the electromotive force driving LSD growth  depends on the difference between the time-integrated kinetic  and  current helicity associated with fluctuations. When the system is helically kinetically forced (KF),  the growth of the large scale helical field is accompanied by growth of small scale magnetic (and current) helicity which ultimately quench the LSD.   Here, using both simulations and theory,  we study the complementary  magnetically forced(MF)  case  in which the system is forced with an electric field that supplies magnetic helicity.  For this MF case,
the kinetic helicity  becomes the back-reactor that saturates the LSD. Simulations of both MF and KF cases can be  approximately modeled with the same equations of magnetic helicity evolution, but with complementary initial conditions.
A key difference between KF and MF cases  is  that  the helical large scale field  in the MF case grows with the same sign of injected magnetic helicity,  whereas  the large and small scale magnetic helicities grow with opposite sign for the KF case.  The  MF case  can arise  even when the thermal pressure is approximately smaller than the magnetic pressure, and requires only that helical small scale magnetic fluctuations
 dominate helical velocity fluctuations in LSD driving. We suggest that LSDs in accretion discs and Babcock models of the solar  dynamo are actually MF LSDs.

\end{abstract}

\section{Introduction}

Understanding the magnetohydrodynamics of large scale magnetic field generation in turbulent astrophysical rotators is the enterprise of large scale dynamo (LSD)  theory.   The LSD problem is usually posed in  kinetically forced (KF) circumstances, whereby an initially weak field is subject to relatively strong hydrodynamic forcing. LSDs then describe the growth or sustenance of fields on spatial or time scales large compared to  the largest  scale of the  underlying  turbulent forcing.   Small scale dynamos (SSD), in contrast, describe field growth on scales at or below the turbulent forcing scale. LSDs and SSDs are contemporaneous  and interacting.  Understanding that interaction, and how KF LSDs saturate have been the subject of much research.

The LSDs of classic 20th century textbooks on the subject (\cite{1978MFge.book.....M}, \cite{1979cMFt.book.....P}, \cite{1980opp..bookR....K}, \cite{1983MFa..book.....Z}) do not predict LSD saturation.  A related  fact  is that these original  LSDs  do not conserve magnetic helicity, the inclusion of which has proven effective at improving the prediction of LSD saturation.
Progress in understanding principles of KF LSD saturation has emerged from  simple numerical experiments to compare with theoretical predictions of  mean field models. The simplest experiments are those of the so called $\alpha^2$ dynamo in which isotropically driven kinetic helicity is injected into a periodic box at intermediate wave number $k=5$ and the $k=1$ large scale field grows (\cite{2001ApJ...550..824B}).
The saturation of the LSD in such simulations is best modeled by theories that solve for the coupled evolution of large scale helical magnetic field and the small scale helical magnetic field. In such models, the electromotive force (EMF, $\bold{\mathcal{E}}$)  of the mean
field theory emerges to be the difference between kinetic helicity and current helicity, each multiplied
by a corresponding correlation time.  This difference was first proposed spectrally in \cite{1975JFM....68..769F}  and derived and solved dynamically as a two-scale mean field theory in  \cite{2002PhRvL..89z5007B}. The results of the dynamical  mean field theory  agree with the saturation seen in the simulation of  \cite{2001ApJ...550..824B}. Recently the three scale version of the theory was also compared to numerical simulations(\cite{2012MNRAS.419..913P}.
Further extensions and approaches for understanding LSDs  via tracking magnetic helicity evolution and flow for more realistic open systems that include boundary flux terms is a fruitful ongoing enterprize.

The basic insight gained from  the basic periodic box KF $\alpha^2$ LSD studies is that as the kinetic helicity is forced and drives large scale  helical magnetic fields, the small scale helical field grows because magnetic helicity is largely conserved. The build up of the small scale current helicity associated with the small scale field then off-sets the  driving kinetic helicity and quenches the LSD.  This physical mechanism and the equations that describe it also motivate consideration of the complementary case of driving the initial system with current helicity rather than kinetic helicity.  This was investigated analytically in \cite{2004PhPl...11.3264B} where it was found that indeed driving with current helicity produces a magnetically forced(MF) analogue to the $\alpha^2$ dynamo. In this paper we present simulations of this MF analogue to the $\alpha^2$ dynamo and compare the results with a  two-scale mean field theory.
We will force the system in the induction equation with an electric field that drives small scale magnetic helicity.  Because the MF system is driven magnetically, there is no ``kinematic'' regime in the sense that the magnetic field is strong from the outset. However the initial velocity fluctuations are weak. Thus the analogue of the kinematic
regime for the MF $\alpha^2$ case is  a ``magnematic'' regime where the small scale velocity fluctuations are small. As we will see, both simulations and theory show that it is indeed the buildup of the kinetic helicity that ends the magnematic regime and quenches the MF analogue of the $\alpha^2$ dynamo.

Simulations of an  inverse cascade of magnetic helicity  was also found in  \cite{2006ApJ...640..335A} but with a focus on non-local  transfer functions in the interpretation of the inverse cascade, and without a dynamical theoretical model in terms of a mean field theory.  Here we will present new simulations, and compare the simulations and theory. We also address the astrophysical relevance of MF LSDs.


In section 2 we more specifically discuss the basic problem to be solved and the computational methods.
In section 3 we present the basic results of the simulations. In section 4 we develop the theory in more detail and discuss the comparison between theory and  simulations.
In section 4 we discuss why  MF LSDs are in fact of basic conceptual relevance to LSDs of accretion discs,
 Babcock type stellar dynamos, and magnetic relaxation of astrophysical coronae and  laboratory
 fusion plasma configurations.  We also discuss some basic open questions.
 We conclude in section 5.


\section{Problem to be studied and Methods}
\begin{figure*}
\centering{
\mbox{%
   \subfigure[$b_r$=0.88]{
     \includegraphics[width=6cm]{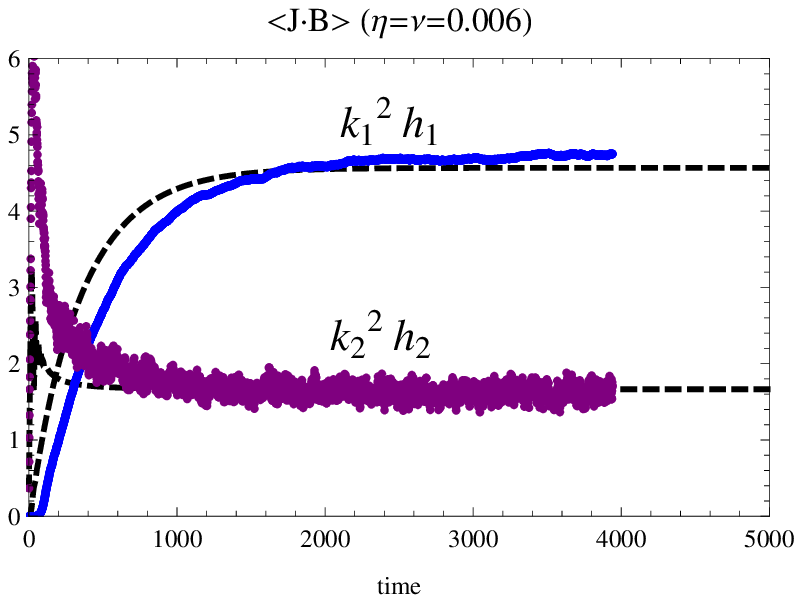}
     \label{h1h2006}
   }\,
   \subfigure[$b_r$=1.5]{
     \includegraphics[width=6cm]{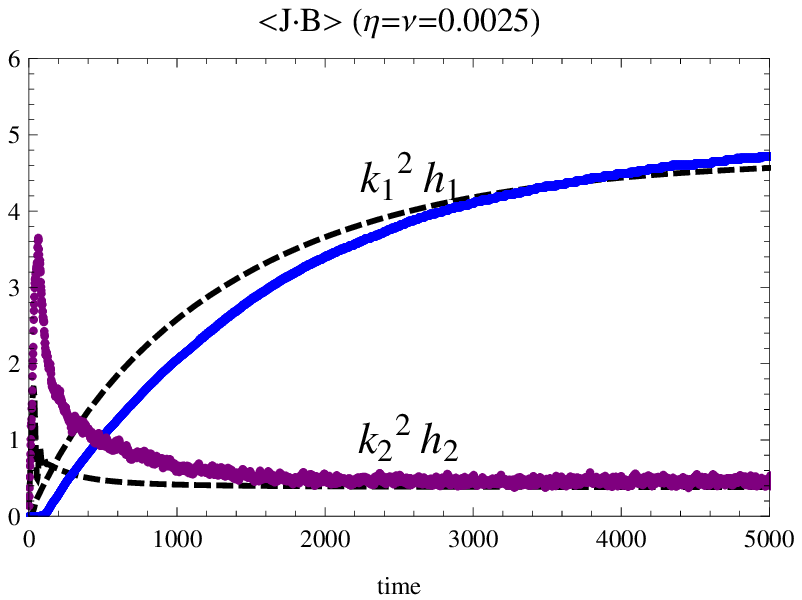}
     \label{h1h20025}}\,
   \subfigure[]{
     \includegraphics[width=6cm]{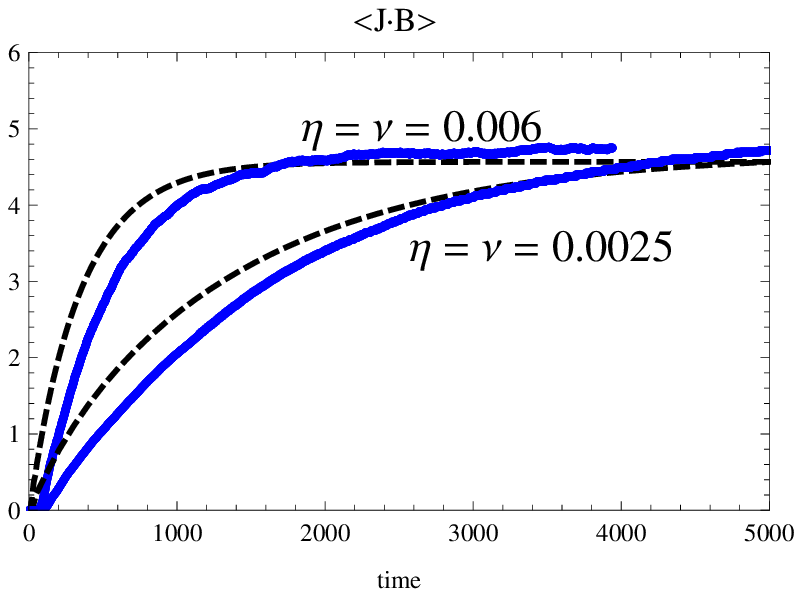}
     \label{h1006and0025}
   }\,
     }
\caption{{\bf (a)} and {\bf (b)}: Time evolution of dimensionless current helicity of large scale field from simulations (blue curve) and small scale field (purple curve) compared
with theoretical predictions (dotted lines)  form our 2-scale model for the large scale
$k_1^2h_1$ and small scale $k_2^2h_2$ current helicities. The first two  panels are for  the two different magnetic diffusivities shown. Current helicity is normalized in units of $b_r^2/k_2$($H_i=h_ib_r^2/k_2$) and
time in units of $k_2b_r$($t=\tau\,k_2b_r$). {\bf (c)} Large scale $k_1^2h_1$ from panels (a) and (b) shown on the same plot.}
\mbox{%
   \subfigure[]{
     \includegraphics[width=8cm]{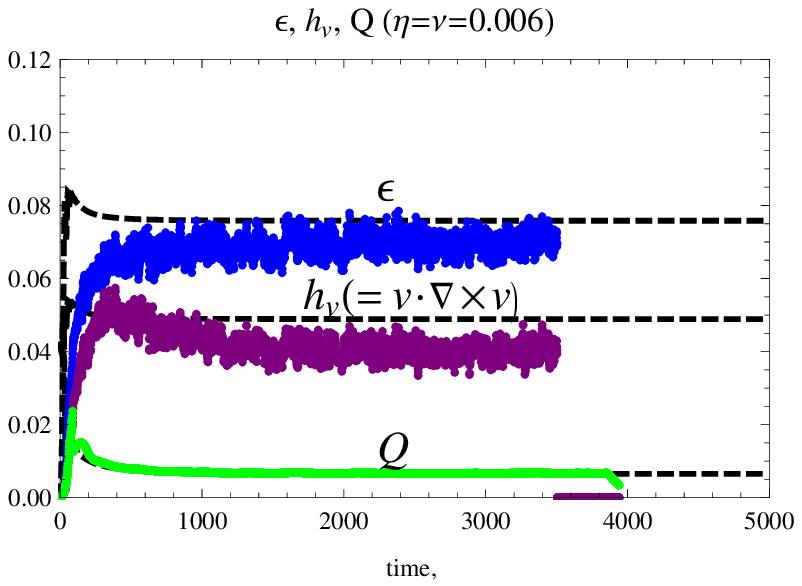}
     \label{eqhv006}
   }\,
   \subfigure[]{
     \includegraphics[width=8cm]{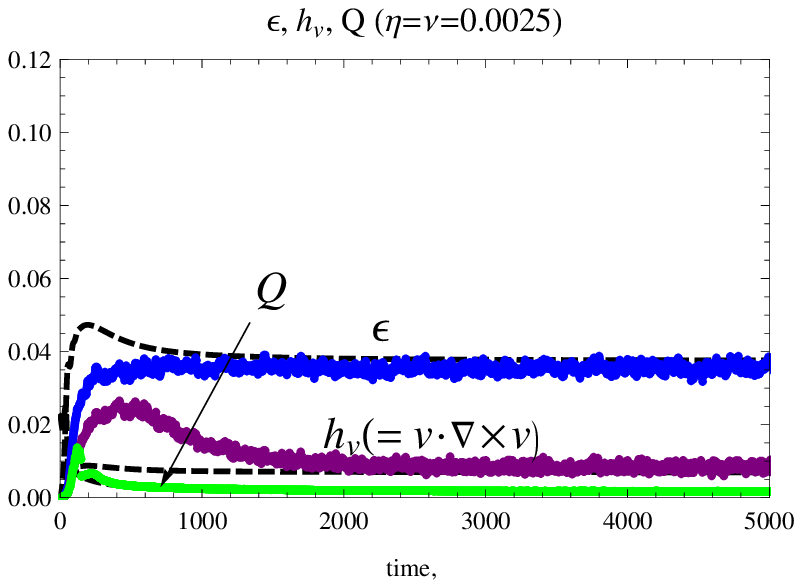}
     \label{eqhv0025}}
}
\caption{{\bf (a)} and {\bf (b)}Dimensionless kinetic energy  ${\bf \varepsilon}\equiv \langle v^2\rangle/\langle b_r^2\rangle$, kinetic helicity  $h_v=\langle {\bf v}\cdot \nabla \times{\bf v}\rangle/(k_2\langle b_r^2\rangle)$ and electromotive force (EMF) : $Q=\bold{\mathcal{E}}_{||}/\langle b_r^2\rangle$, for two different magnetic diffusivities.}}
\end{figure*}
We used the high-order finite difference Pencil Code(\cite{2001ApJ...550..824B}) along with the message passing interface(MPI) for parallelization.
The equations solved by the code  are
 \begin{eqnarray}
\frac{D \rho}{Dt}&=&-\rho \nabla \cdot {\bf u}, \\
\frac{D {\bf u}}{Dt}&=&-c_s^2\nabla \mathrm{log}\, \rho + \frac{{\bf J}\times {\bf B}}{\rho} \nonumber \\
& & + \frac{\rho_0 \nu}{\rho}(\nabla^2 {\bf u}+\frac{1}{3}\nabla \nabla \cdot {\bf u}),\\
\frac{\partial {\bf A}}{\partial t}&=&{\bf u}\times {\bf B} -\eta\,{\bf J} + {\bf f}.
\label{MHD equations in the code}
\end{eqnarray}
where $\rho$ is the density, $\bf u$ is the velocity, $\bf B$ is the magnetic field, ${\bf J}$ is the current density,  $D/Dt=\partial / \partial t + {\bf u} \cdot \nabla$, is the advective derivative, `$\eta$' is magnetic diffusivity, `$\nu$' is viscosity, and $c_s$  is the sound speed, and $\bf f$ is  forcing function. The default units of the pencil code are `$cgs$'.

We employ a periodic box of  dimensionless spatial volume $(2 \pi)^3$  and a mesh size of $216^3$.
No large scale velocity forcing is imposed. Instead, our forcing function $f$ is placed in the magnetic induction equation and is a fully helical, Gaussian random force-free function($\nabla\times{\bf f}\propto{\bf f}$) given by  ${\bf f}(x,t)=N\,{\bf f_k}(t)\, exp\,[i\,{\bf k}(t)\cdot {\bf x}+i\phi(t)]$, where $N$ is a normalization factor and ${\bf k}(t)$ is the forcing wave number with $|{\bf k}(t)|\sim 5$ ($4.5<|{\bf k}(t)|<5.5$), and $|\bf f_k|$=0.03($|\bf f_k|$ of 0.07 was used in (\cite{2012MNRAS.419..913P})).
This forcing function can be thought of as imposing a small scale external electric field $\bfe_{ext}$
such that  $\bf f = {\bf e}_{ext}$. It is a source term for magnetic
 helicity for the small scale field proportional to  $\langle \bfe_{ext}\cdot \bfb\rangle$.

From the solutions of the equations, we compute a range of quantities including the  box averaged spectra of kinetic energy($E_k$), magnetic energy($E_M$),  kinetic helicity($\langle {\bf v}\cdot \nabla \times {\bf v}\rangle$) and magnetic helicity($\langle {\bf A}\cdot {\bf B}\rangle $). From these
quantities we further construct mean quantities for comparisons with a two-scale theoretical
model as discussed in subsequent sections.   We note that although our system is magnetically driven, the ``plasma beta'' (i.e. $\beta_p\equiv $ ratio of magnetic energy density to  thermal energy density) satisfies $\beta_p \gtrsim 1$ most of the time in our simulations.

We will consider Prandtl numbers of unity in this paper, for two different values of the resistivity.  In some of the analysis of our equations we need to integrate over all wavenumbers and it will be noted that in both cases we take $k_{max}$ to be
107. For the two values of $\eta$ that we choose, namely $\eta=0.0025$ and $\eta=0.006$ we find that  $k_{dis}\sim 107$ for $\eta=0.006$ and
$k_{max} \le k_{dis}$ for $\eta=0.0025$ but that very little power in any of the relevant
quantities appears above $k=107$ for either case. Thus we typically used the same upper bound in $k$ for both cases.

For the data analysis, we used Interactive Data Language(IDL).

\section{Simulation Results}

\subsection{Basic Results: Growth of Large Scale Field and Saturation}

Fig.\ref{h1h2006}-\ref{eqhv0025} show a comparison between our simulation results  and
two-scale theory (discussed further in section 3) for the dimensionless large and small scale current helicities  ($k_1^2 h_1$ and $k_2^2h_2$ respectively), kinetic energy ($\epsilon$), kinetic helicity ($h_v$), and
EMF ($\equiv\bold{\mathcal{E}}=\langle \bf v\times \bf b \rangle \sim Q$).  The   simulation data for the large scale field correspond to quantities at $k_1=1$
and the simulation data for the small scale field correspond  to the sum of quantities for  $2<k <k_{max}$.
The analytic two scale model (discussed later in more detail) takes $k_1=1$ for the large scale and  the forcing scale $k_2=5$ for the small
scale.

Roughly,  the  large scale  current helicity  growth in  in Fig. 1 resembles that of  the current  helicity studied in  the KF case (\cite{2012MNRAS.419..913P}).  Unlike the KF case,
the the large scale magnetic (and current) helicity has the same sign as that on the small scale for the present MF case. This is because our forcing in the induction equation injects magnetic helicity of a fixed sign into the system and magnetic helicity is  conserved until resistive terms kick in.
The MF  $\alpha^2 $ LSD thus acts as a dynamical relaxation process: the injection of small magnetic helicity drives the system away from equilibrium and the LSD competes with this injection by converting much of the small scale injected magnetic helicity into large scale magnetic helicity of the same sign.  The magnetic energy of a fixed amount of magnetic helicity is less on large scales than on small scales and much of the energy lost in the relaxation goes  into kinetic energy. Note from Fig.\ref{eqhv006} and \ref{eqhv0025} that a sizeable fraction of the kinetic energy gained is helical.

As we will see in the next subsection, most of  kinetic helicity is on the forcing and smaller scale. The growing kinetic helicity ultimate quenches the LSD in the non-linear regime.
This mechanism of quenching is supported by a comparison between $k_2^2h_2$ curves in Fig.\ref{h1h2006} and \ref{h1h20025} with $h_v$ curves in Fig.\ref{eqhv006} and \ref{eqhv0025}. In dimensionless form, the LSD of section 4 is driven by the difference between $h_2$ and $h_v$. When these two are roughly equal (the difference being only equal to smaller diffusion and resistive terms) significant quenching is expected. Dividing the $k_2^2 h_2$ curves by
$k_2^2=25$ shows indeed that the value $h_2\sim h_v$ in the late time regimes of the plots.
In this context, note that Fig 1c shows that the higher Reynolds number  (equal to magnetic Reynolds  in this paper) case takes longer for the large scale field to saturate.  This is analogous to the KF case where the higher
magnetic Reynolds number case also takes longer to saturate.  However for the MF case, it is the viscosity not the resistivity that damps the back reaction because the back reactor is kinetic helicity.
In contrast,  the KF case saturates because the small scale current helicity builds up, which is damped by
resistivity.  The analytic dependence on magnetic Prandtl number for the MF was discussed
in \cite{2004PhPl...11.3264B}.
We do not pursue that further here.



\subsection{Time Evolution of Energy and Helicity Spectra}

\begin{figure*}
\centering
\mbox{%
   \subfigure[]{
     \includegraphics[width=6cm]{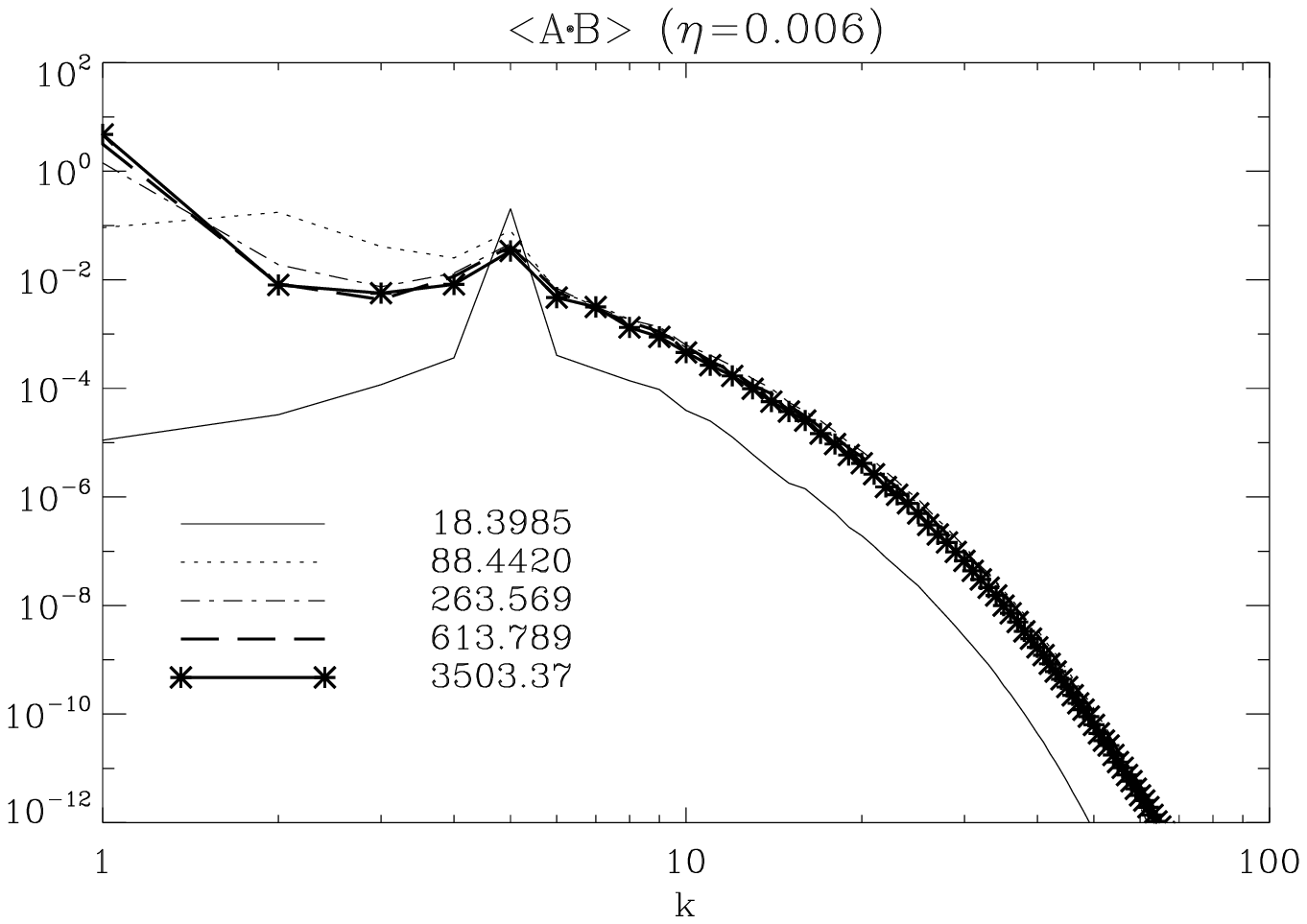}
     \label{magnetichelicity006}
   }\,
   \subfigure[]{
     \includegraphics[width=6cm]{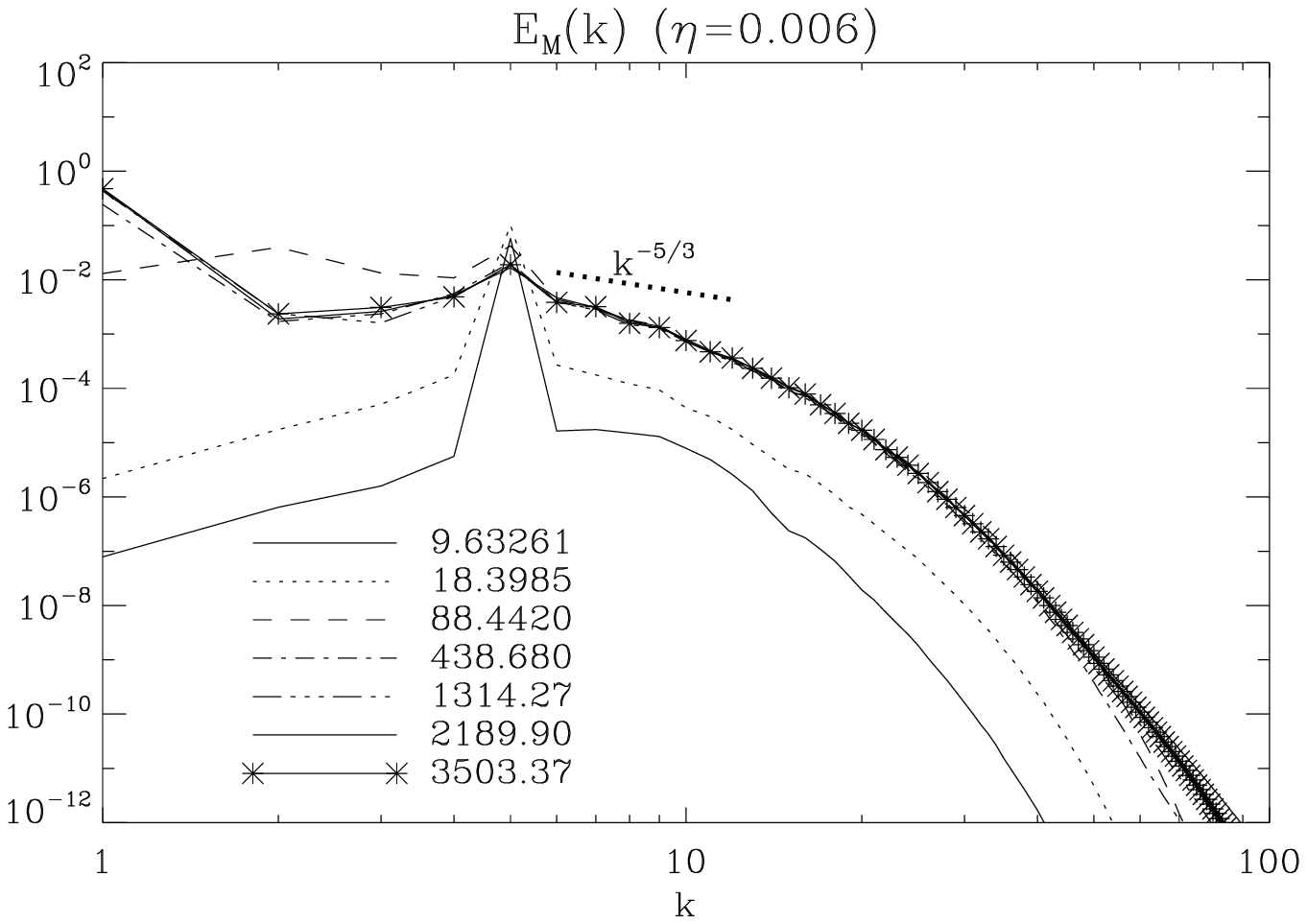}
     \label{magneticenergy006}
   }\,
   \subfigure[]{
     \includegraphics[width=6cm]{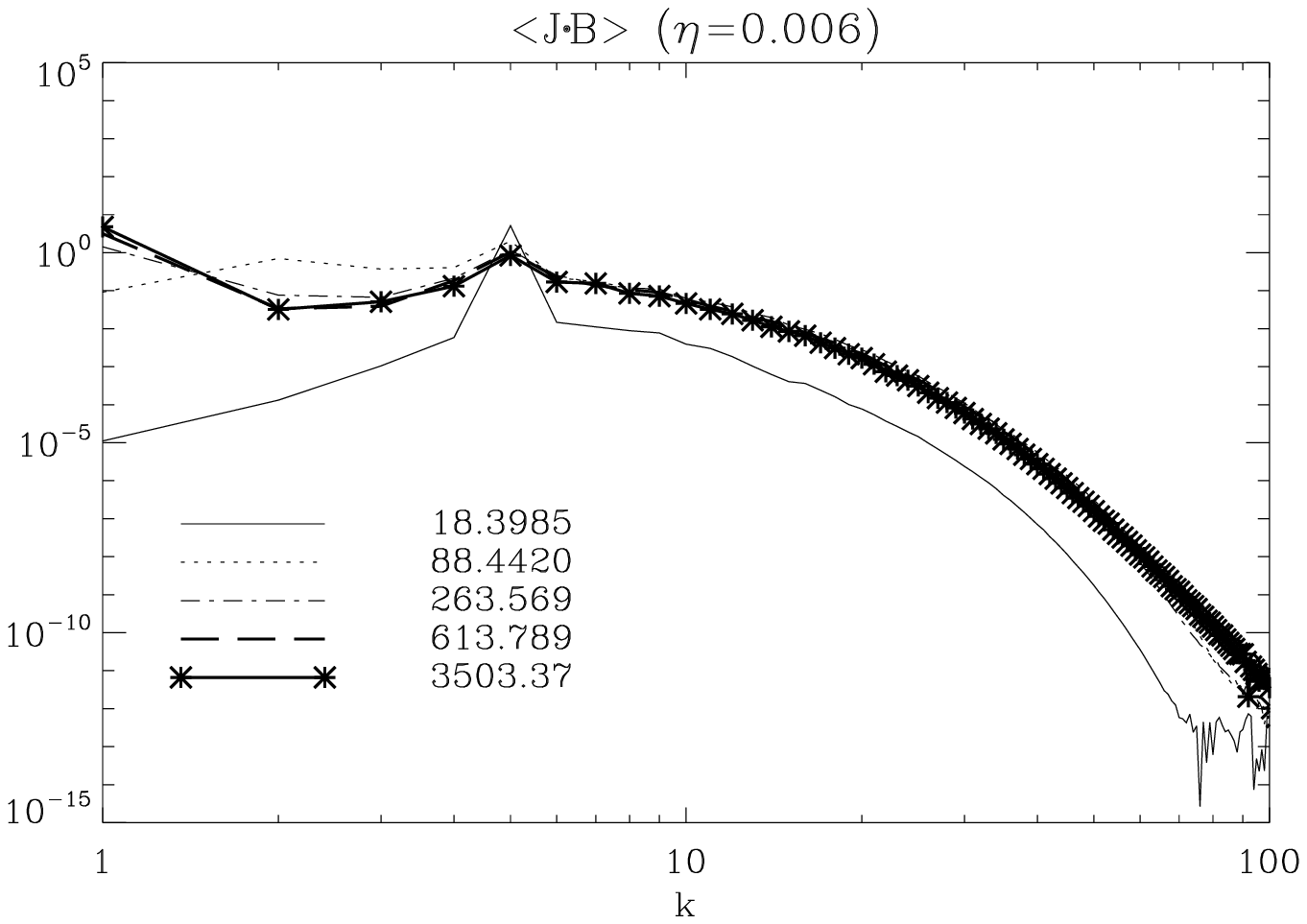}
     \label{currenthelicity006}}
}
\caption{Time evolution for the spectra of  (a) magnetic helicity (b) magnetic energy (c) current helicity for the $\eta= 0.006$ run.}
\mbox{%
   \subfigure[]{
     \includegraphics[width=6cm]{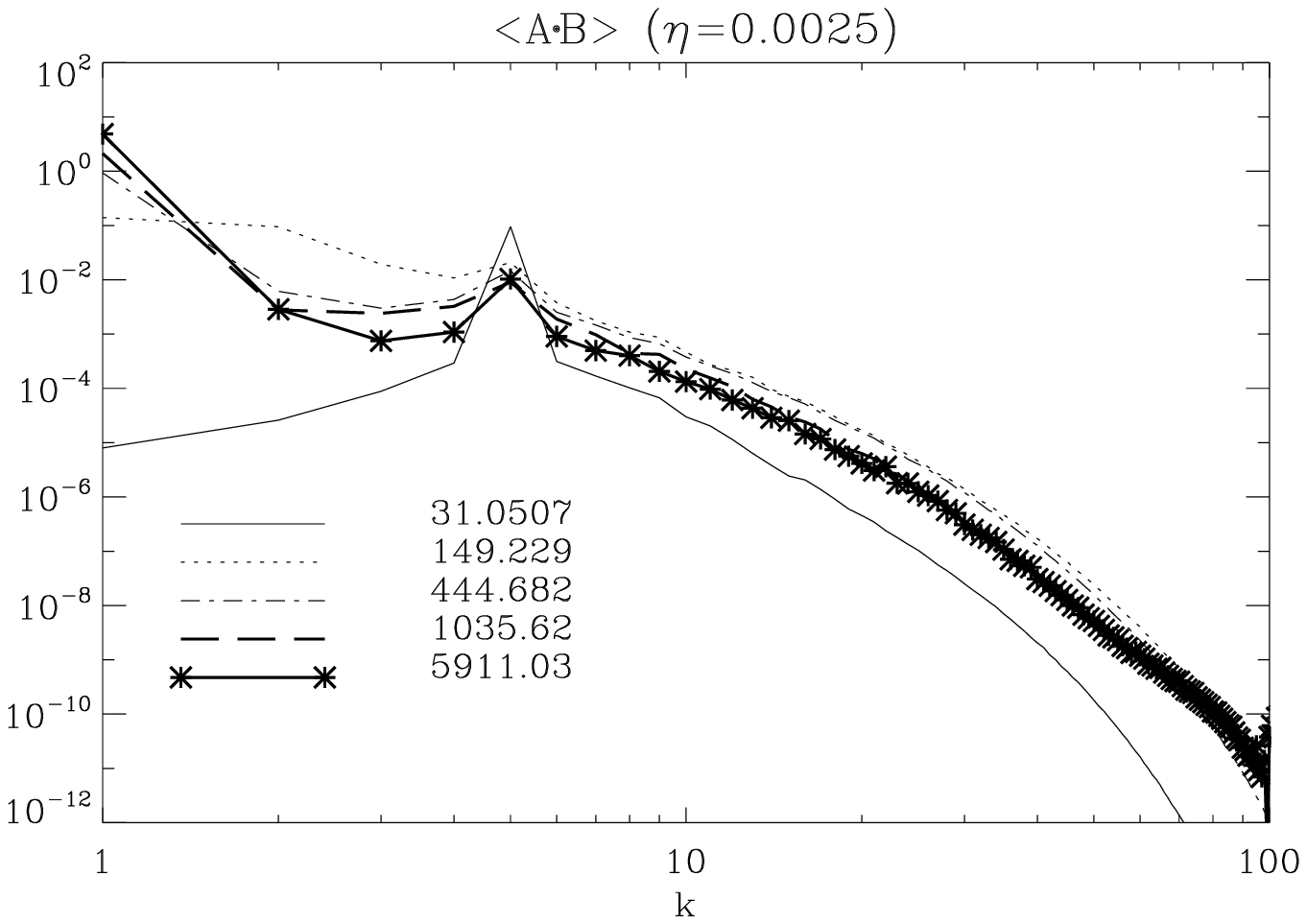}
     \label{magnetichelicity0025}
   }\,
      \subfigure[]{
     \includegraphics[width=6cm]{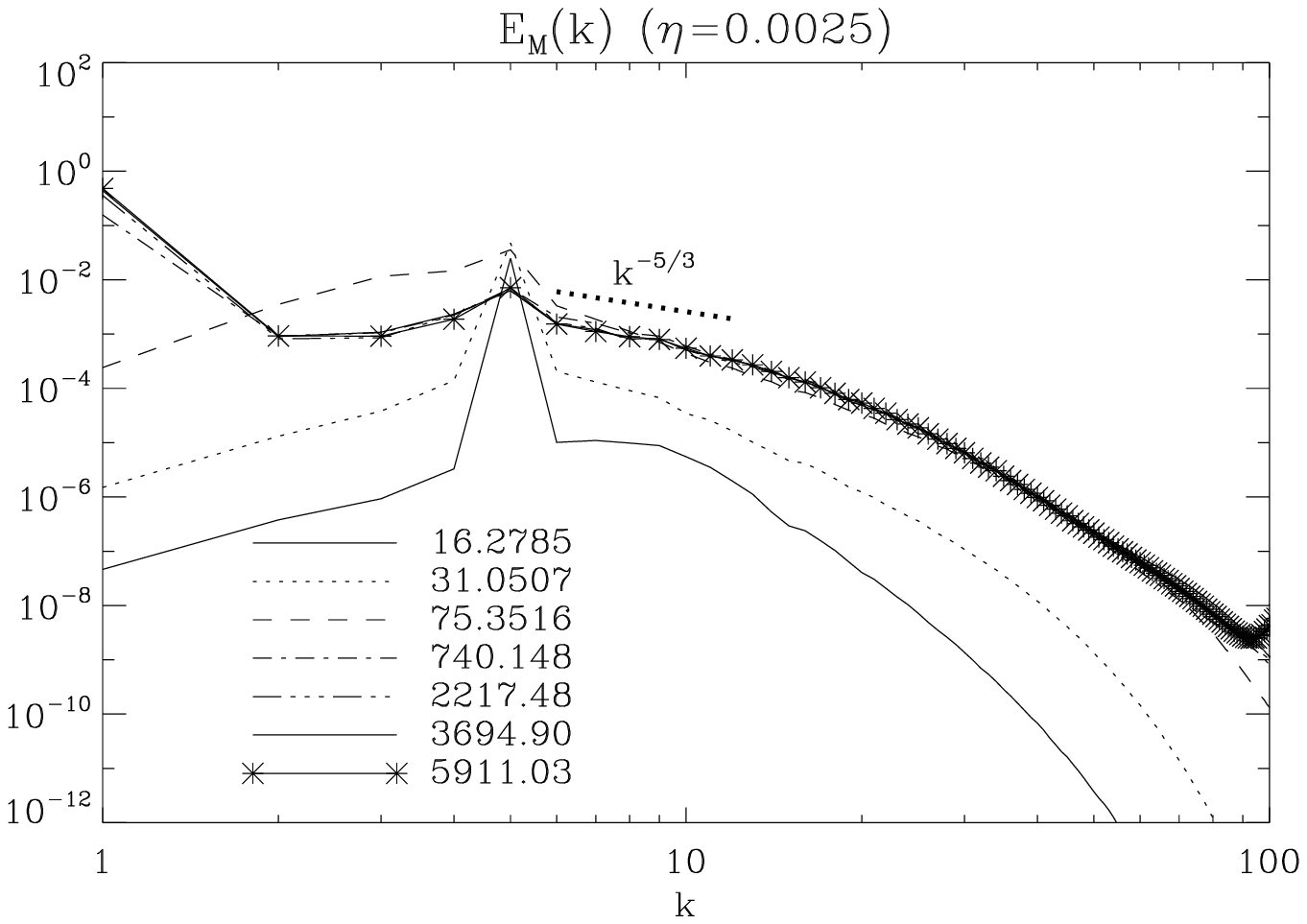}
     \label{magneticenergy0025}}\,
   \subfigure[]{
     \includegraphics[width=6cm]{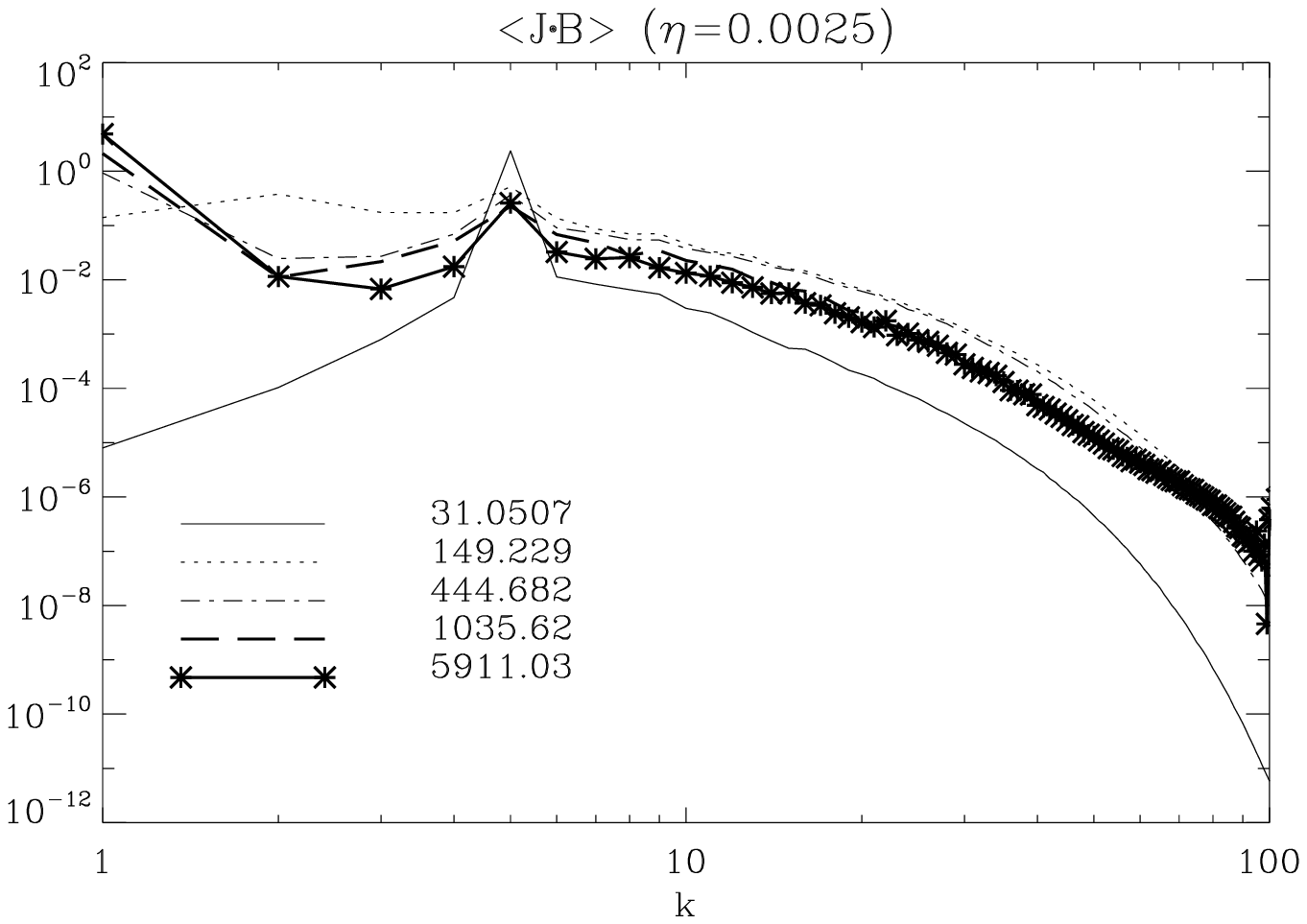}
     \label{currenthelicity0025}}
}
\caption{Time evolution for the spectra of  (a) magnetic helicity (b) magnetic energy (c) current helicity for the $\eta= 0.0025$ run.}

\end{figure*}

For the MF case, Figs.\ref{magnetichelicity006}-\ref{currenthelicity0025}
show the time evolution of the magnetic helicity, magnetic energy and current helicity
spectra as the simulation  for the two different values of diffusivities used. Note the  strong
early peak at the forcing scale and the ultimate transfers to the $k=1$ scale.

\begin{figure*}
\centering
\mbox{%
   \subfigure[]{
     \includegraphics[width=8cm]{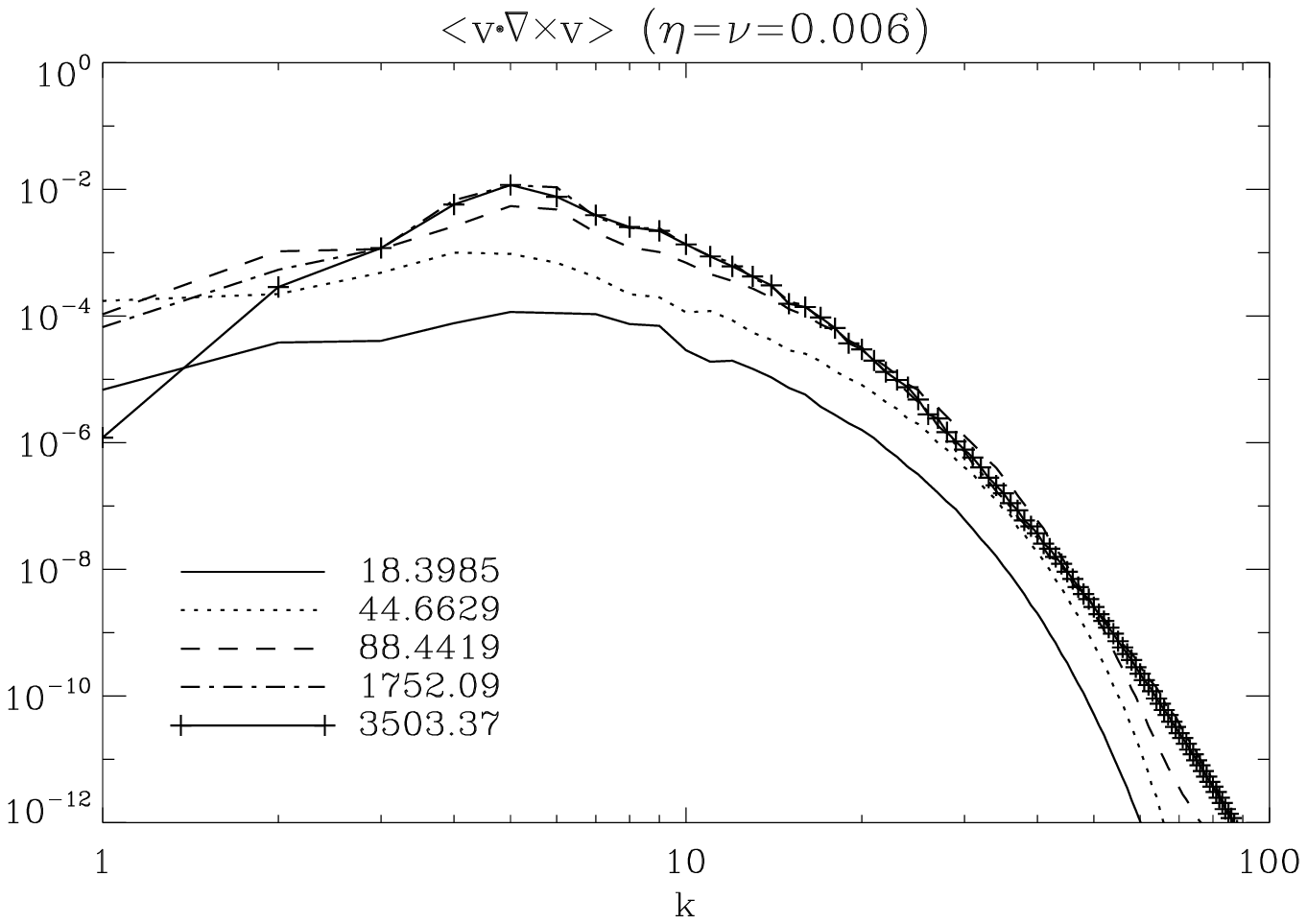}
     \label{kinetichelicity006magneticforcing}
   }\,
   \subfigure[]{
     \includegraphics[width=8cm]{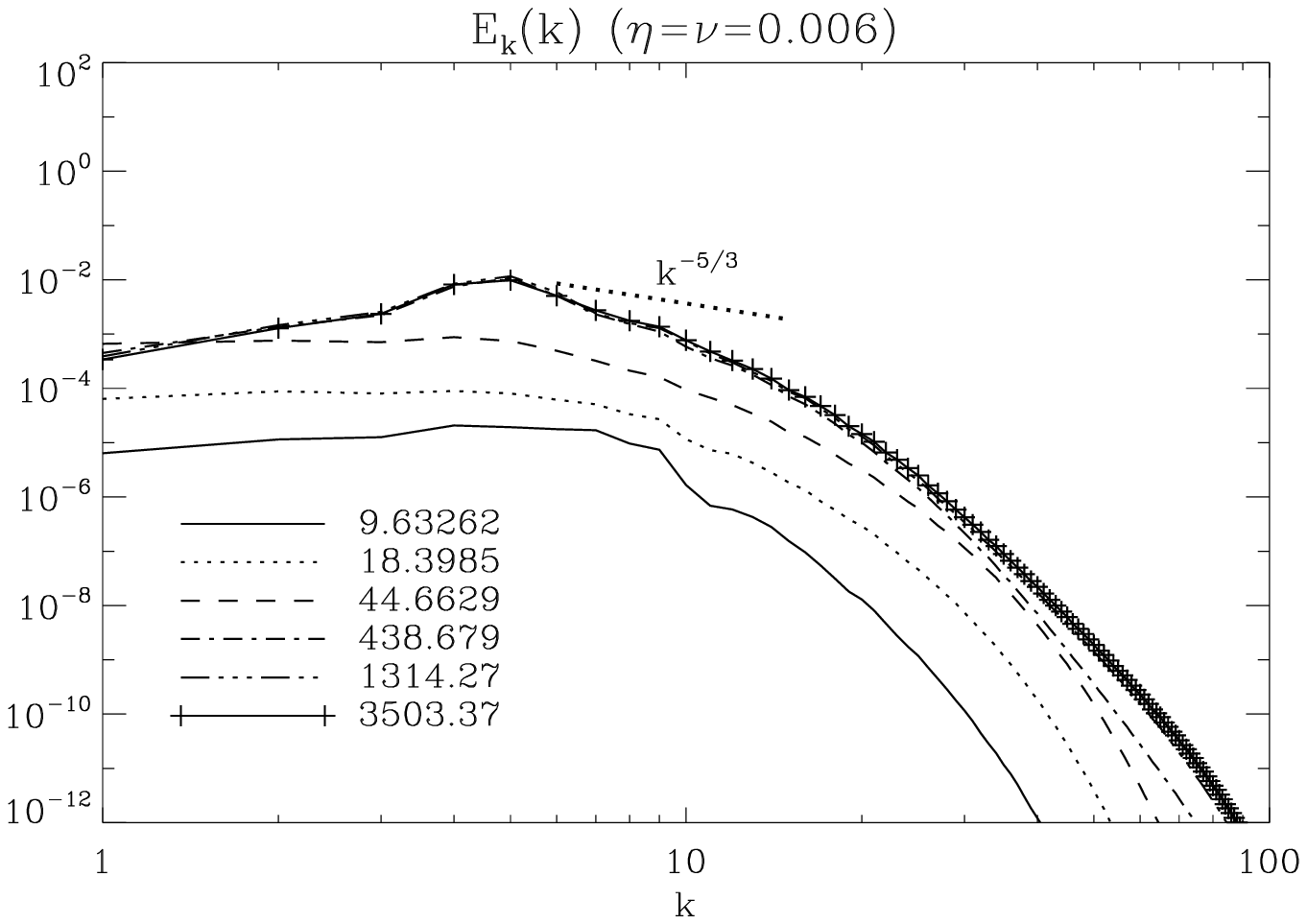}
     \label{kineticenergy006magneticforcing}
   }
}
\caption{Time evolution for the spectra of  (a) kinetic helicity and (b) kinetic  energy
for the $\eta= 0.006$ run.}

\mbox{%
   \subfigure[]{
     \includegraphics[width=8cm]{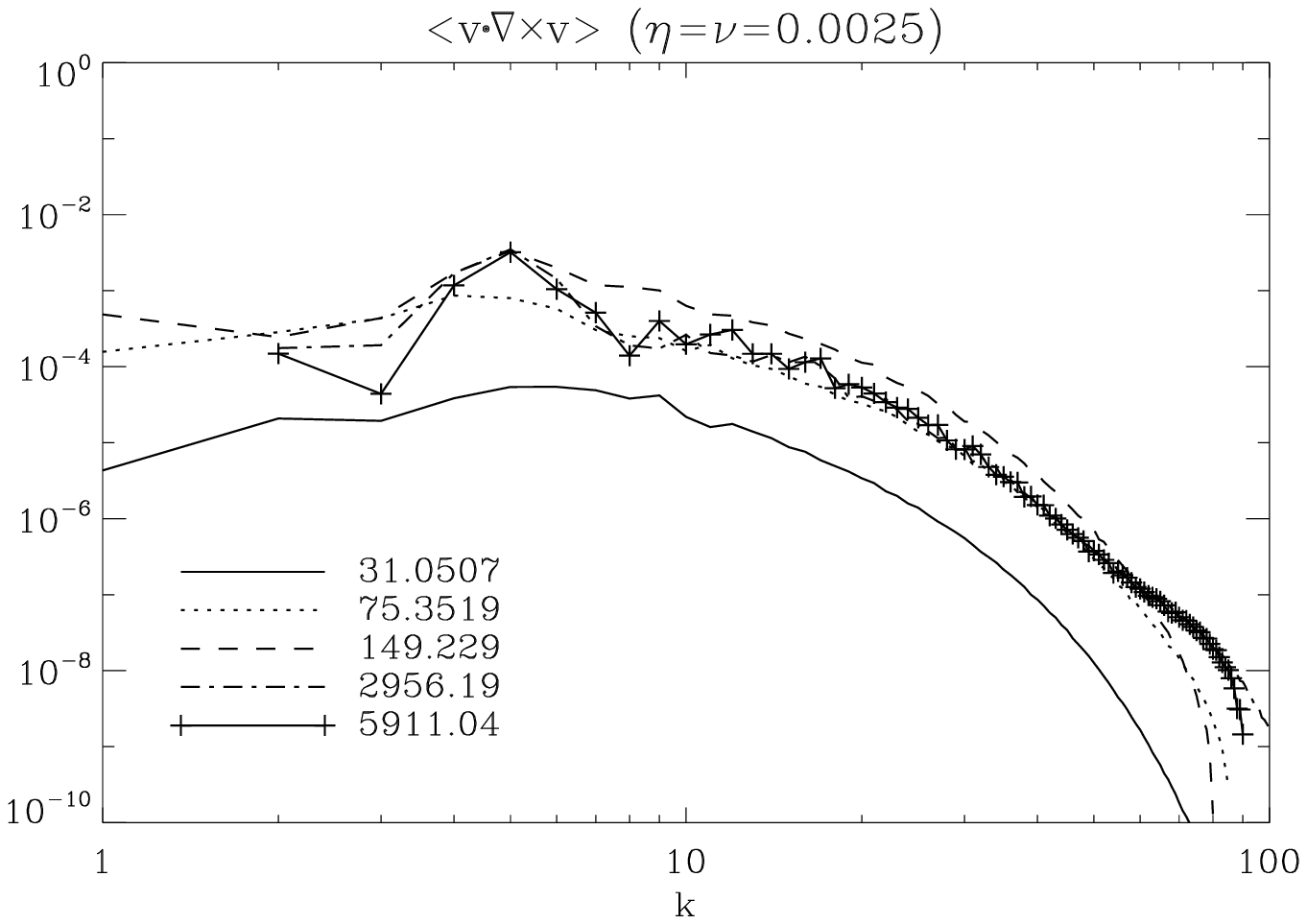}
     \label{kinetichelicity0025magneticforcing}
   }\,
      \subfigure[]{
     \includegraphics[width=8cm]{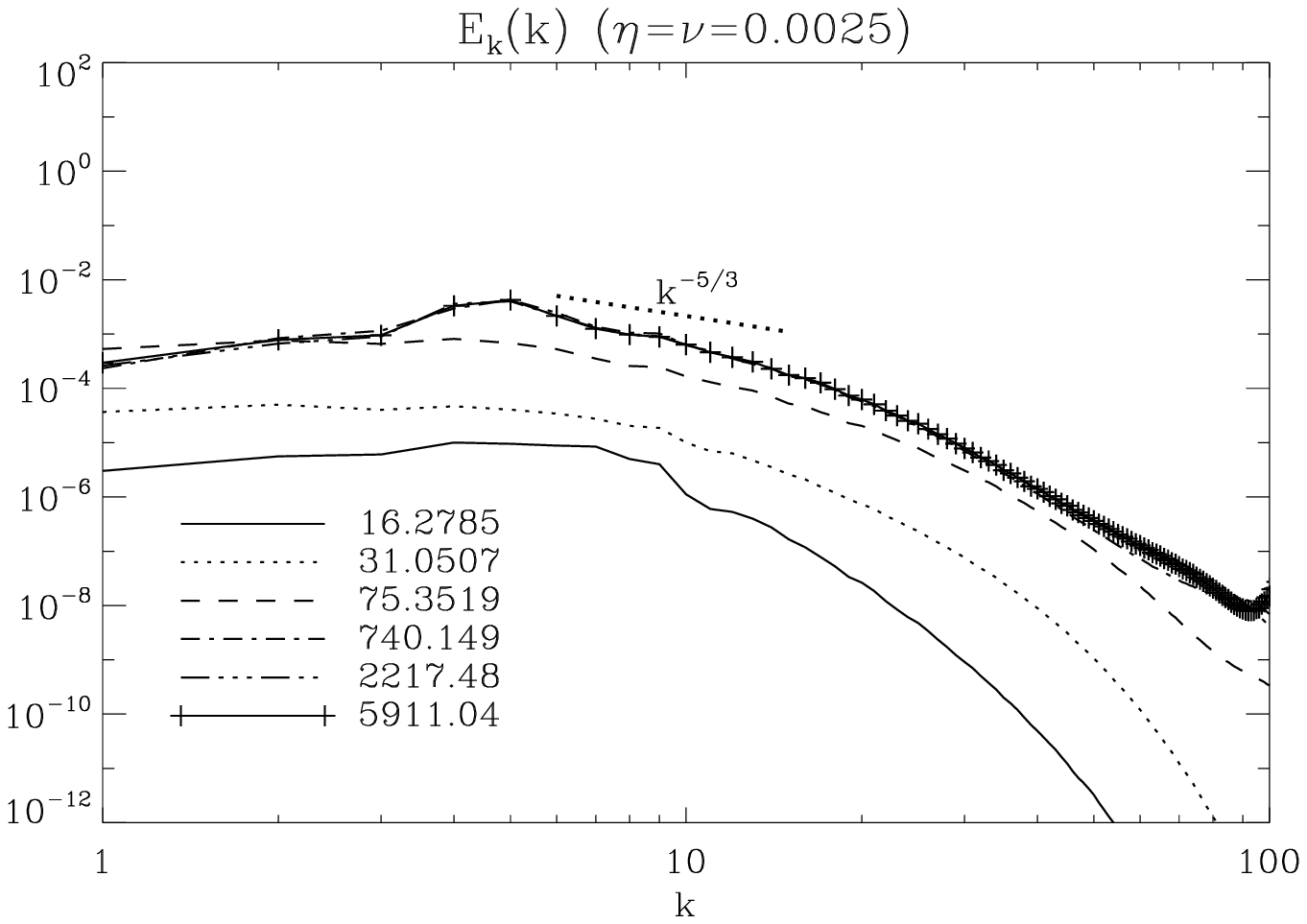}
     \label{kineticenergy0025magneticforcing}
                  }
}
\caption{Time evolution for the spectra of  (a) kinetic helicity and (b) kinetic  energy  for the $\eta= 0.006$ run.}

\end{figure*}

Figs.\ref{kinetichelicity006magneticforcing}-\ref{kineticenergy0025magneticforcing}
show the corresponding spectral evolution of the kinetic helicity and kinetic energy.
Note that the kinetic energy is less peaked that than the magnetic energy at the forcing scale
for this MF case and that there is some growth of kinetic energy at the large $k=1$ scale.
As we discuss later, this is due to the Lorentz force at $k=1$ before the large scale field becomes force free.

\begin{figure*}
\centering
\mbox{%
   \subfigure[]{
     \includegraphics[width=8cm]{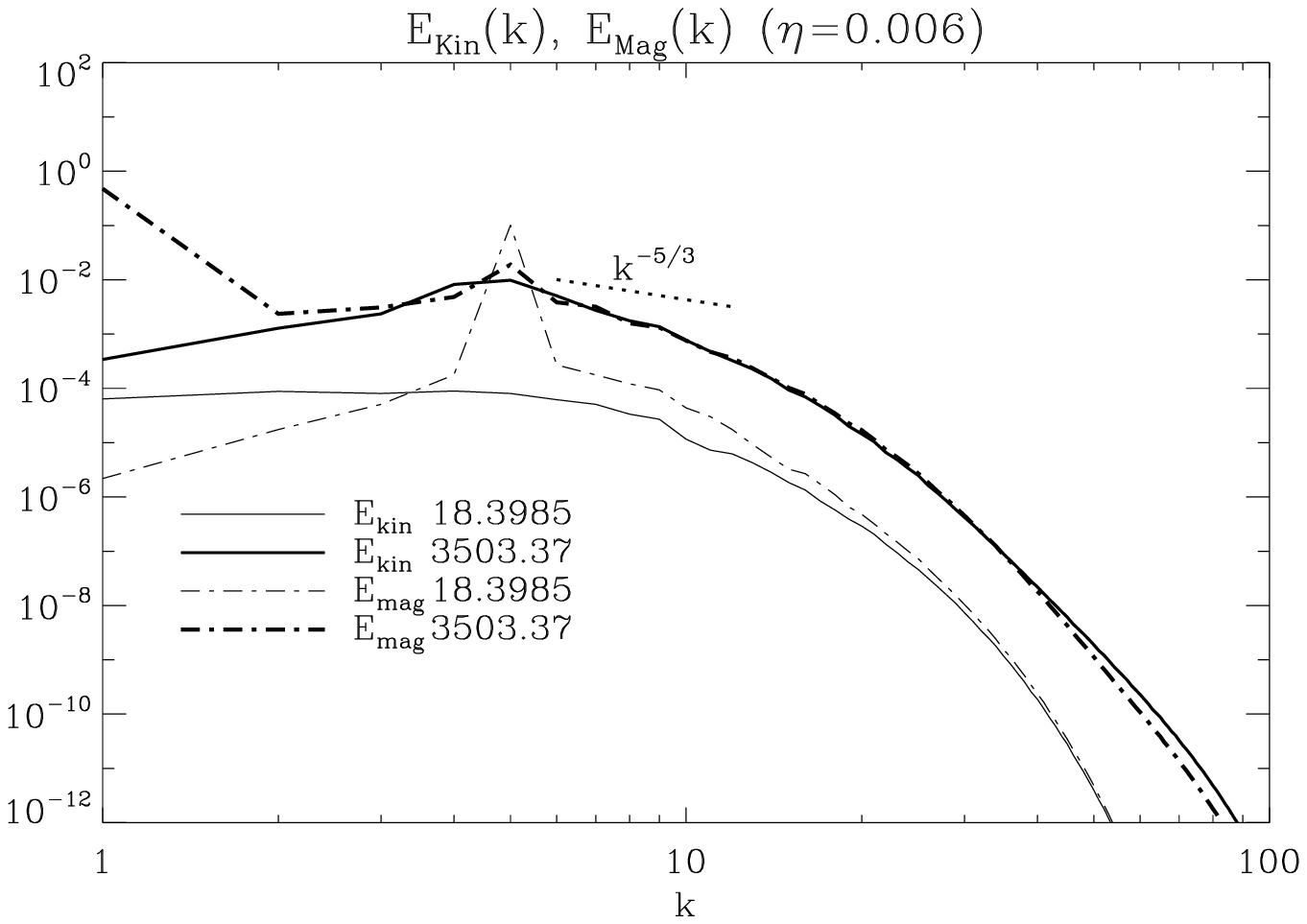}
     \label{kineticmagneticenergy006magneticforcing}
   }\,
   \subfigure[]{
     \includegraphics[width=8cm]{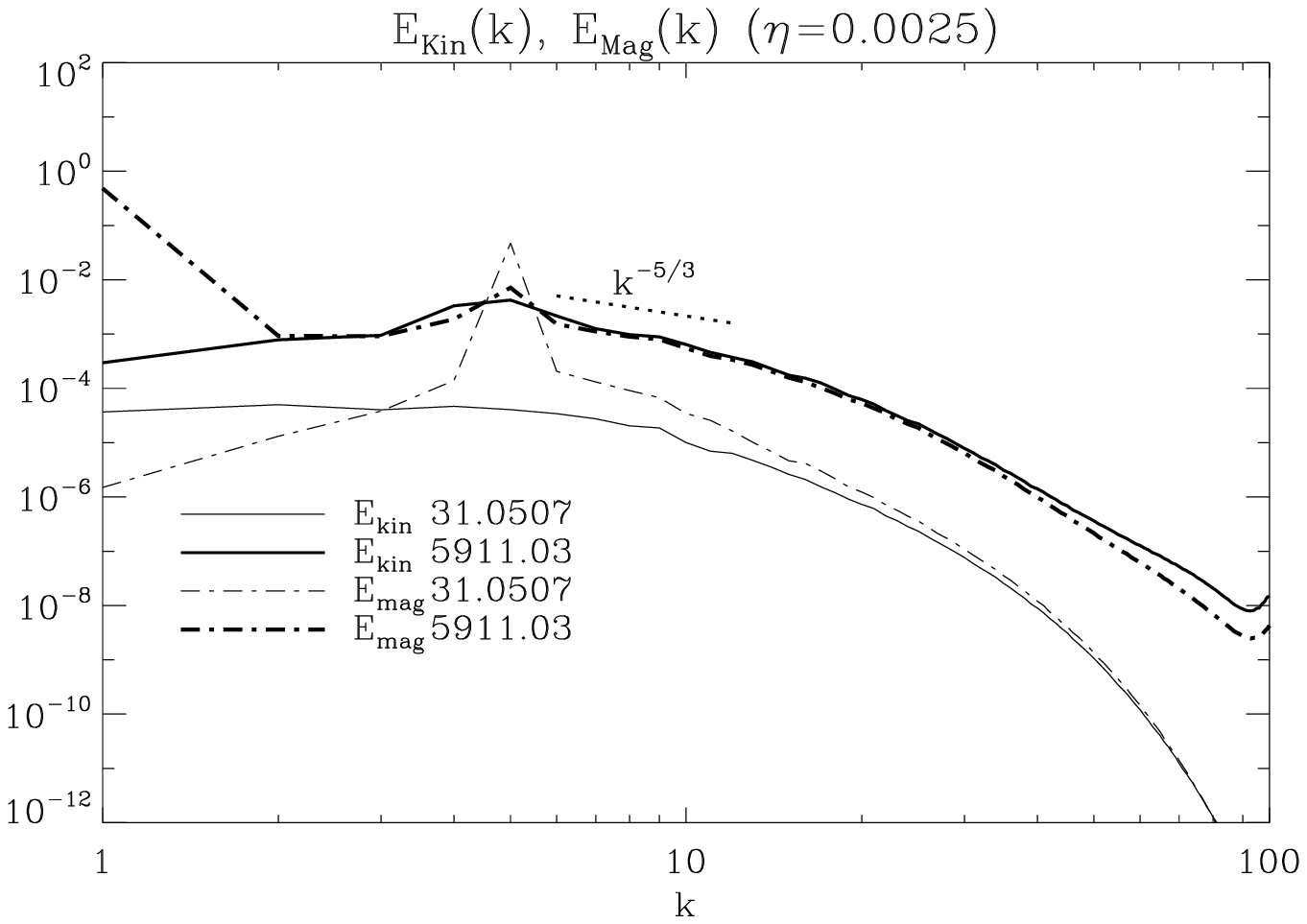}
     \label{kineticmagneticenergy0025magneticforcing}
   }
}
\caption{Comparison of time evolving spectra of  $E_{kin}$ and $E_{mag}$ for the $\eta =0.006$ and $\eta=0.0025$ runs.  Note that the large scale $k=1$
kinetic energy remains small  whereas the large scale magnetic  field grows..}

\end{figure*}

\begin{figure*}
\centering
\mbox{
   \subfigure[]{
     \includegraphics[width=10cm]{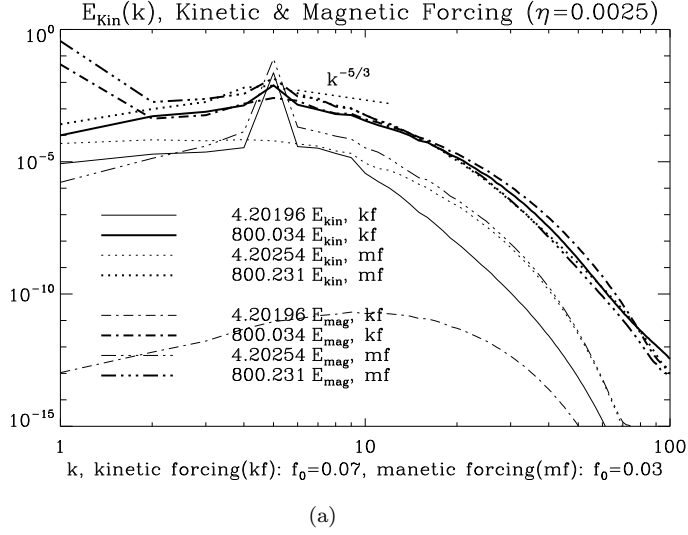}
     \label{kineticmagneticenergy0025kineticmagneticforcing}
   }
}
\caption{Time evolution of  the $E_{kin}$ and $E_{mag}$  spectra for the MF case of $\eta =0.0025$   plotted together with that of the KF case  \citep{2012MNRAS.419..913P}.}

\end{figure*}

Fig.\ref{kineticmagneticenergy006magneticforcing} and \ref{kineticmagneticenergy0025magneticforcing} compare the kinetic and magnetic energy spectra on the same plots and Figure  \ref{kineticmagneticenergy0025kineticmagneticforcing} shows the time evolution of the kinetic and magnetic energy  spectra for the MF case of the present paper compared with the KF simulations of \cite{2012MNRAS.419..913P}. Note that the KF case shows a stronger
peak at the forcing scale for the kinetic energy than the MF case, and the MF case shows a stronger peak
at the forcing scale in magnetic energy than the KF case.

















To study the spectral evolution of various quantities and their migration between scales,
it is useful to define  average wave numbers  with the different bases defined below (\cite{2004tise.book.....D}, \cite{2012MNRAS.419..913P}) as seen below:
\begin{eqnarray}\label{kavab}
\langle k\rangle_{\bf A\cdot \bf B}
=\frac{\int_{k_i}^{k_{max}} k {\bf A} \cdot {\bf B}\, dk}{\int_{k_i}^{k_{max}} {\bf A} \cdot {\bf B}\, dk}\simeq \frac{\sum_{k_i}^{k_{max}} k \bf A\cdot \bf B}{\sum_{k_i}^{k_{max}}  \bf A\cdot \bf B},
\end{eqnarray}
\begin{eqnarray}\label{kavbb}
\langle k\rangle_{E_{mag}}=\frac{\int_{k_i}^{k_{max}} k {\bf B} \cdot {\bf B}\, dk}{\int_{k_i}^{k_{max}} {\bf B} \cdot {\bf B}\, dk}\simeq \frac{\sum_{k_i}^{k_{max}} k \bf B\cdot \bf B}{\sum_{k_i}^{k_{max}} \bf B\cdot \bf B},
\end{eqnarray}
\begin{eqnarray}\label{kavjb}
\langle k\rangle_{\bf J\cdot \bf B}=\frac{\int_{k_i}^{k_{max}} k {\bf J}\cdot {\bf B}\, dk}{\int_{k_i}^{k_{max}} {\bf J}\cdot {\bf B}\, dk}\simeq \frac{\sum_{k_i}^{k_{max}} k^3 \bf A\cdot \bf B}{\sum_{k_i}^{k_{max}} k^2 \bf A\cdot \bf B},
\end{eqnarray}
\\
where  $k_i$ will be taken  either as $k_i=1$ or $k_i=2$ depending on whether we compute the averages
for all wave numbers or extract the large scale $k=1$ in order to compute  small scale averages.

Analogously, for $\langle k \rangle_{{\bf v} \cdot {\bf v}}$, $\langle k \rangle_{{\bf v} \cdot {\bf \omega}}$, and $\langle k \rangle_{{\bf \omega} \cdot {\bf \omega}}$
\begin{eqnarray}\label{kavvv}
\langle k\rangle_{\bf v\cdot \bf v}
=\frac{\int_{k_i}^{k_{max}} k {\bf v} \cdot {\bf v}\, dk}{\int_{k_i}^{k_{max}} {\bf v} \cdot {\bf v}\, dk}\simeq \frac{\sum_{k_i}^{k_{max}} k \bf v\cdot \bf v}{\sum_{k_i}^{k_{max}}  \bf v\cdot \bf v},
\end{eqnarray}
\begin{eqnarray}\label{kavvw}
\langle k\rangle_{\bf v \cdot \bf \omega}=\frac{\int_{k_i}^{k_{max}} k {\bf v} \cdot {\bf \omega}\, dk}{\int_{k_i}^{k_{max}} {\bf v} \cdot {\bf \omega}\, dk}\simeq \frac{\sum_{k_i}^{k_{max}} k \bf v\cdot \bf \omega}{\sum_{k_i}^{k_{max}} \bf v\cdot \bf \omega},
\end{eqnarray}
and
\begin{eqnarray}\label{kavww}
\langle k\rangle_{\bf \omega\cdot \bf \omega}=\frac{\int_{k_i}^{k_{max}} k {\bf \omega}\cdot {\bf \omega}\, dk}{\int_{k_i}^{k_{max}} {\bf \omega}\cdot {\bf \omega}\, dk}\simeq \frac{\sum_{k_i}^{k_{max}} k \bf \omega\cdot \bf \omega}{\sum_{k_i}^{k_{max}} \bf \omega\cdot \bf \omega}.
\end{eqnarray}

The time evolution of the average wave numbers in the kinetic bases
is shown in  Fig.\ref{kvvvwww006} and  Fig.\ref{kvvvwww0025} for the MF case, which can be compared with the KF case (\cite{2012MNRAS.419..913P})
of  Fig.\ref{kvvvwww006kf} and Fig.\ref{kvvvwww0025kf}. The
time evolution of the average wave numbers in the magnetic bases for the MF case is  shown in Fig.\ref{kabbbjb006} and Fig.\ref{kabbbjb0025} which can be compared with KF case in Fig.\ref{kabbbjb006kf} and Fig.\ref{kabbbjb0025kf}).
Each of the figures  shows two kinds of averages the ``total average''
for which the integration range extends from  $k_i=1$  to $k_{max}=107$ or the
``small scale'' for which the integration range extends  from $k_i=2$ to $k_{max}=107$.
To distinguish these, we use the subscript ``tot''
to indicate  averages defined by the former and ``s'' by the latter. For example, we
write $k_{\jbt}$ or $k_{\jbs}$, or $k_{tot}$ and $k_{s}$ when not specifying the basis.





Using $k=1$ for the large scale and the range $k=[2, 107]$ for the small scale,
Fig.\ref{emag006early} and \ref{jb006early} highlight the increase of small scale magnetic energy
and current  helicity at early times that results from the MF case followed by the longer term growth of the large scale field.  Eventually the magnetic energy is dominated by that on the  large scale.
 Fig \ref{eklargesmall006} and Fig.\ref{vwlargesmall006} show the analogous curve for the
 kinetic energy and kinetic helicity.  Although the kinetic energy and kinetic helicity also grow
 in response to the MF, the kinetic energy is dominated by small scale contributions.

Fig.\ref{kineticenergy006magneticforcing} and \ref{kineticenergy0025magneticforcing} show that most of the kinetic energy remains on small scales during the simulations  but there is some growth of large scale kinetic energy.  There is no independent inverse cascade of  velocity in 3-D, only in 2-D (\cite{2004tise.book.....D}) so the growth of any large scale velocity  in our case is expected to be the direct result of the Lorentz force on those scales. Indeed according to Fig.\ref{kvvvwww006}, \ref{kvvvwww0025} we can see that the average wavenumbers
$k_{\vet}$ and $k_\ves$ are unequal, highlighting that there is some energy in the large scale $k=1$ velocity. This inequality disappears as time passes and  coincidence occurs approximately when the magnetic field becomes fully helical(Fig.\ref{fhmfhk006}, \ref{fhmfhk0025}, \ref{fhmfhklargescale006}, \ref{fhmfhklargescale0025}). Since the magnetic field can only transfer energy to
velocity fields when  the   Lorentz force(${\bf J}\times {\bf B}$, Eq.(\ref{Navier Stokes equation})) is finite on this scale,
no energy can be transferred to the kinetic eddy  once the field becomes fully helical.
  The evolution of the helicity fractions of both the kinetic and magnetic helicity   fractions as a function of wave number and time are shown in
  Figs \ref{fhmfhk006} and \ref{fhmfhk0025}. The time evolution of the  helicity
  fractions for $k=1$ are shown in Fig.\ref{fhmfhklargescale006} and  \ref{fhmfhklargescale0025}).


The evolution of the mean wavenumbers seen in
Fig.\ref{kabbbjb006} and \ref{kabbbjb0025} highlights the evolution of the
magnetic energy and helicity spectra in mf case.   At very early times $k_\jbt=k_\bbt=k_\abt=k_\jbs=k_\bbs=k_\abs=5$, i.e., $k_{tot}$$\sim$$k_{s}$ in all bases. This means the magnetic energy resides almost exclusively at the forcing scale($k_f$) in this early time regime. From Eq.(\ref{kavab}), the average $k$ at this state is simply $k{\bf A}\cdot {\bf B}/{\bf A}\cdot {\bf B}|_{k=k_f}$, which is `$5$'. Since magnetic field and current helicity are helical(${\bf J} = k {\bf B}=k^2 {\bf A}$), the basis independent profile in the early time is explained.

By $t\sim 50$, the   energy begins to cascade both forward (to small scales) and inversely (to large scales).   For  $\sim50< t <\sim100$,  we still have  $k_{tot}$$\sim$$k_{s}$ in all bases (since all 6 curves of
of e.g. Fig.\ref{kabbbjb006}) continue to overlap in this time regime) but
$\langle k \rangle_{\bf A \cdot \bf B}$, $\langle k \rangle_{\bf B \cdot \bf B}$, and $\langle k \rangle_{\bf J \cdot \bf B}$ split.  The  fact that the quantities all decrease together in this regime shows that although there magnetic energy is still overwhelmingly contained in the small scales,   the quantities are evolving toward larger scales but they have not crossed over to the large scale yet. The two regimes just discussed can be referred to as   ``pre-LSD I''  and ``pre-LSD II'' regimes because there is little growth of magnetic energy at $k=1$.

Beyond $t\sim 100$ Fig.\ref{kabbbjb006} and \ref{kabbbjb0025} show the influence of the LSD.
Note that  the evolution of $k_s$ and $k_{tot}$ for each quantity now  deviate from each other.  That is,  calculating the $k$ averages is dramatically affected by changing the lower integration bound from $k=1$ to $k=2$. By this time, $\bfb$, $\bfv$, and EMF have grown and  driven  growth of  the large scale $k=1$ field.  Also at this time, the sign of   $dk_s/dt$ changes from negative to positive. For example, $\langle k\rangle_{\bf A \cdot \bf B, s}$ of $\eta=\nu=0.0025$ (Fig.\ref{kabbbjb0025}),
transitions from negative to positive   growth over the regime $\sim100 < t < \sim200$.
The reason for this change is two fold: (i) The LSD takes  inversely transferred magnetic helicity from the  small scales to $k=1$,  thus dropping magnetic helicity and magnetic energy  out of the bins that contribute to $k_s$(Eq.\ref{ratio of ktot and ks1}),  and (ii) The small helicity is continuously driven at the forcing scale $k=5$.



 The buildup of  kinetic helicity eventually slows the LSD. Already around $t\sim200$ the
 slopes of the curves in Fig.\ref{kabbbjb006} and \ref{kabbbjb0025}
 flatten, seemingly consistent with the time at which
$\langle {\bf v}\cdot \nabla \times  {\bf v}\rangle$ approaches saturation and approximately equals
$\langle{\bf j} \cdot {\bf b}\rangle$. That marks  the end of the kinematic regime,
 as discussed in the previous subsection.  Eventually, the system reaches a steady-state.

\subsubsection{Relation between $k_{tot}$ and $k_s$}

We can mathematically relate $k_{tot}$ and $k_s$  for each basis. For example, for the case of $k_\abt$ and $k_\abs$
   (we let $\sum \equiv \sum^{k_{max}}_{k=2}$ for convenience in this subsection),
 we have
\begin{eqnarray}\label{application of ktot and ks1}
&&k_\abt \equiv\frac{\sum k\, {\bf a} \cdot {\bf b} + 1\,\bf \overline{A}\cdot \bf \overline{B}}{\sum {\bf a} \cdot {\bf b} + \bf \overline{A}\cdot \bf \overline{B}}, \quad
k_\abs \equiv\frac{\sum k\, {\bf a} \cdot {\bf b}}{\sum {\bf a} \cdot {\bf b}}\nonumber\\
&&\Rightarrow
k_\abs \sum {\bf a} \cdot {\bf b} + \overline{{\bf A}}\cdot \overline{{\bf B}}
=k_{\abt}\big(\sum {\bf a} \cdot {\bf b} + \overline{{\bf A}}\cdot \overline{{\bf B}}   \big)\nonumber\\
&&\Rightarrow (k_\abs-k_{\abt})\sum {\bf a} \cdot {\bf b}=(k_{\abt}-1)\overline{A}\cdot \bf \overline{B}\nonumber\\
&&\Rightarrow \frac{\overline{{\bf A}}\cdot \overline{{\bf B}}}{\sum {\bf a} \cdot {\bf b}}=\frac{k_\abs-k_{\abt}}{k_{\abt}-1}.
\end{eqnarray}
When $k_\abs$=$k_{\abt}$,
the numerator on the right hand side(RHS) is 0. This means most of
the magnetic  helicity is in the small scale. In contrast, when $k_{\abt}\sim 1$, most of the magnetic  helicity is in the large scale. Generally the time rate of change of the large scale over small scale helicities  is given by
\begin{eqnarray}\label{ratio of ktot and ks1}
\frac{d}{dt}\bigg(\frac{\overline{{\bf A}}\cdot \overline{{\bf B}}}{\sum {\bf a} \cdot {\bf b}}\bigg)
={(\dot{k}_\abs-\dot{k}_{{\bf A} \cdot {\bf B},  tot})(k_{\abt}-1)\over (k_{\abt}-1)^2}\nonumber\\
-{(k_\abs-k_{\abt})\,\dot{k}_{\abt} \over (k_{\abt}-1)^2}.
\label{application of ktot and ks2}
\end{eqnarray}

In the early time regimes I and II, $k_\abs=k_{\abt}$ and $\dot{k}_{\abt}=\dot{k}_\abs$, so `$d\, (\overline{{\bf A}}\cdot \overline{{\bf B}} / \sum {\bf a} \cdot {\bf b})/dt$' is zero. Once $k_\abs\neq k_{\abt}$ and $\dot{k}_{\abt}<\dot{k}_\abs$), this quantity becomes positive, i.e., the inverse cascade sets in.


The equations of this subsection can also be applied to $k_{\jbt}$ and $k_{\jbs}$,
and can be used to quantify the shape of the curves in e.g.
Fig.\ref{kabbbjb006} and \ref{kabbbjb0025}. For example, when $\dot{k}_{\abt} < 0$ and $\dot{k}_{\abs}=0$, the inverse cascade begins to accelerate.


\section{Theoretical Two Scale Model}
\subsection{Basic Equations}
\begin{figure*}
\centering
\mbox{%
   \subfigure[]{
     \includegraphics[width=7cm]{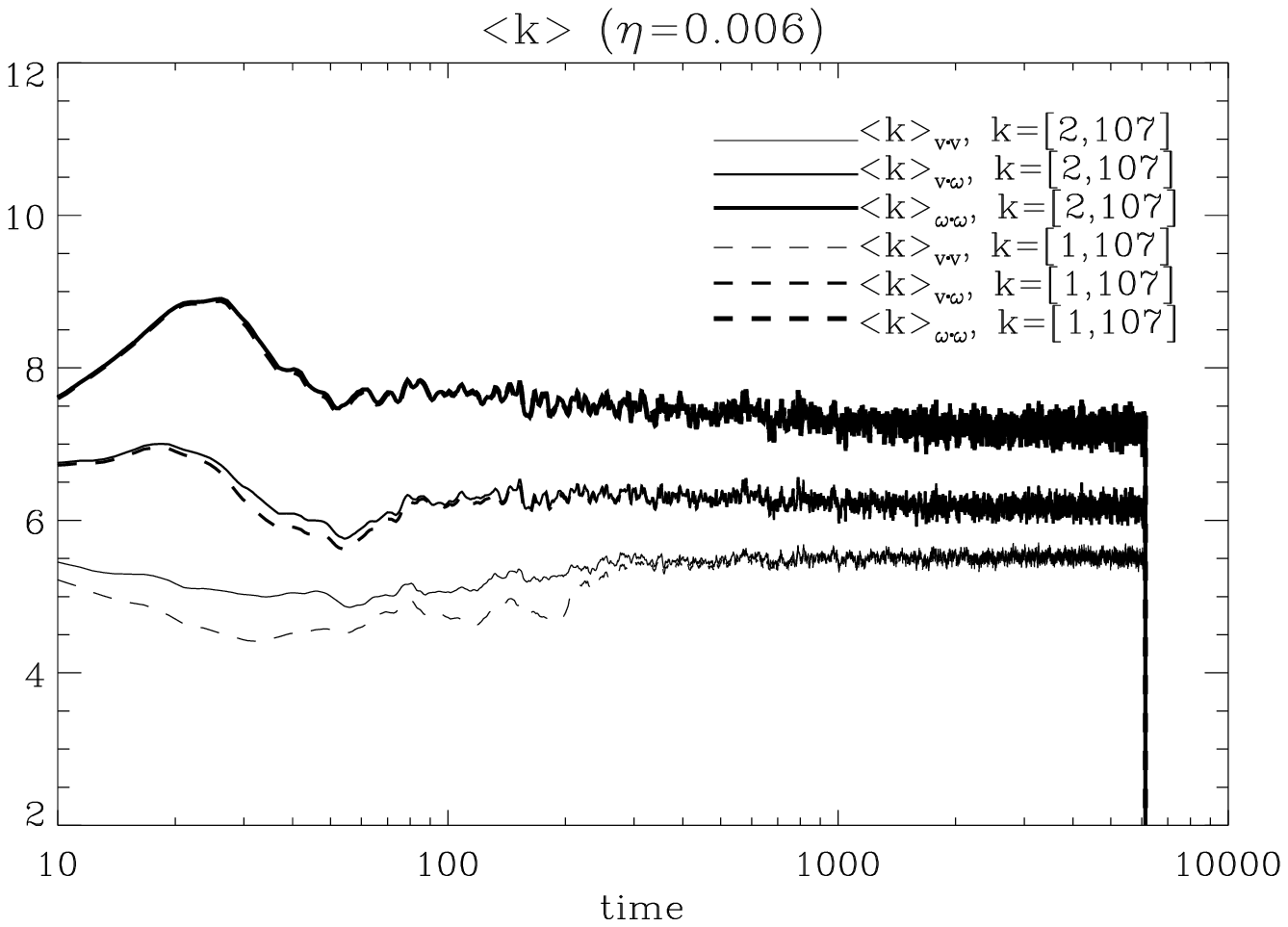}
     \label{kvvvwww006}
   }\,
   \subfigure[]{
     \includegraphics[width=7cm]{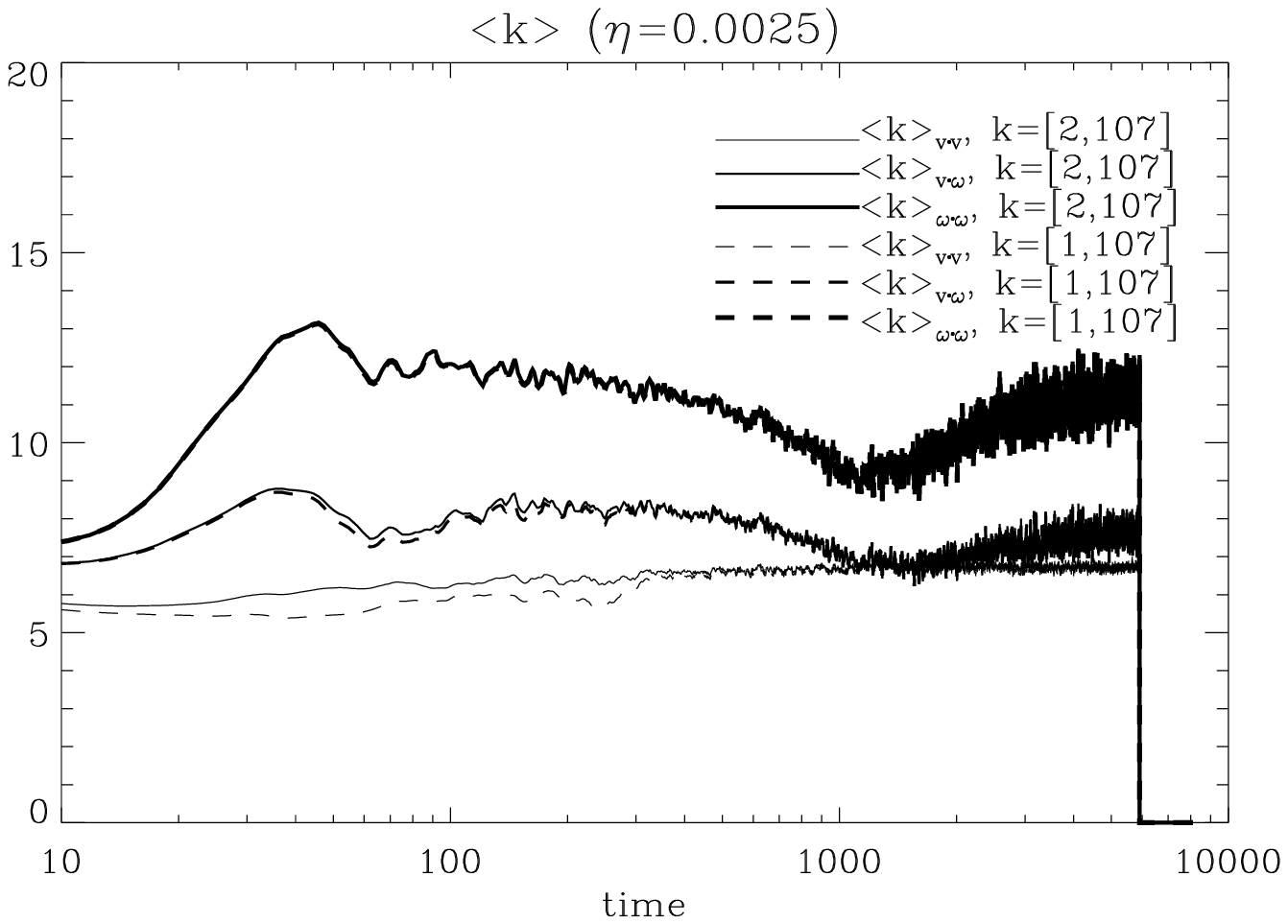}
     \label{kvvvwww0025}
   }
}
\caption{Mean kinetic wavenumbers in the different kinetic bases for the two different
resistivity cases for  MF case.  Simulations used a forcing parameter $f_0$=0.03}.
\mbox{%
   \subfigure[]{
     \includegraphics[width=7cm]{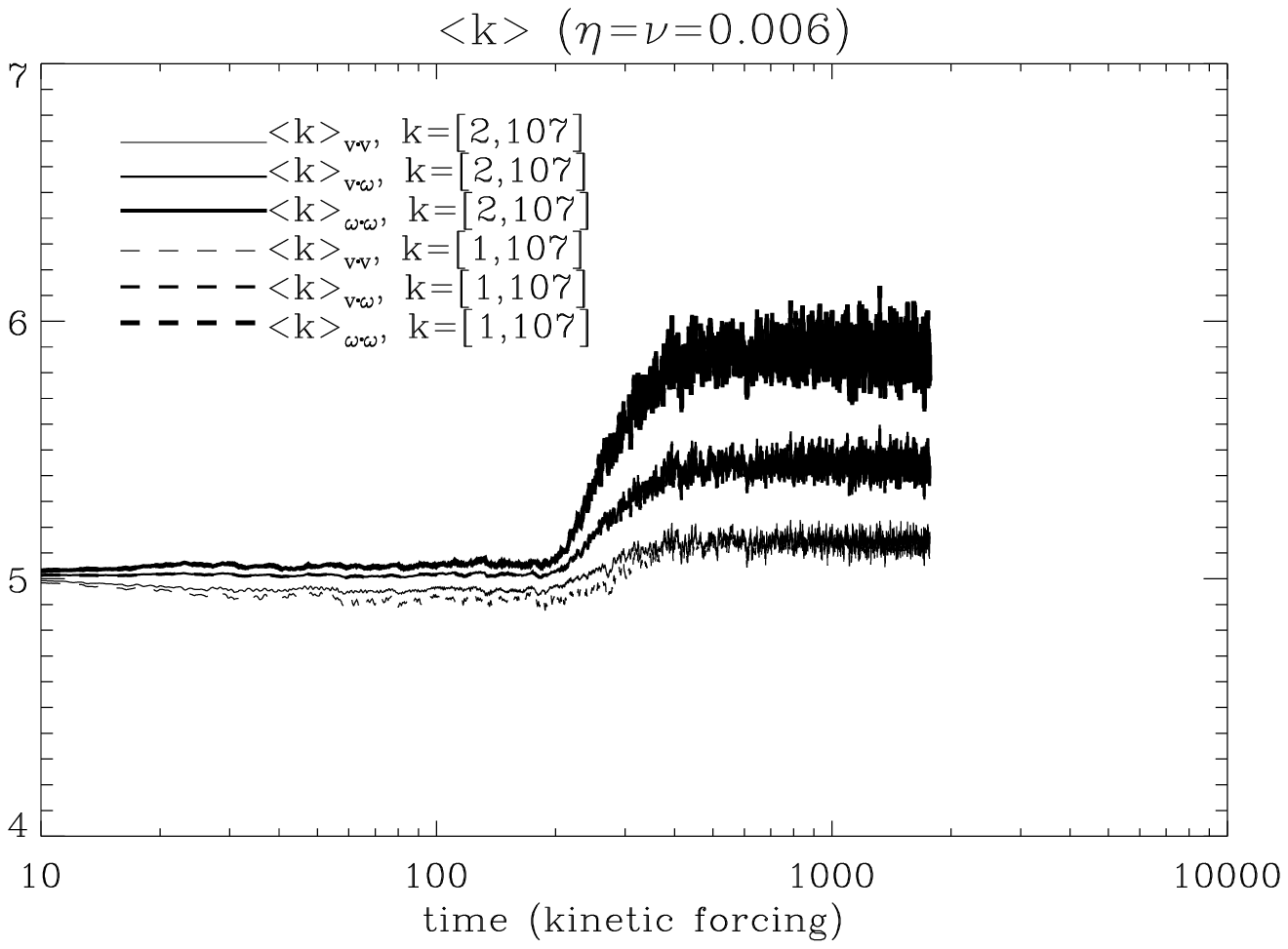}
     \label{kvvvwww006kf}
   }\,
   \subfigure[]{
     \includegraphics[width=7cm]{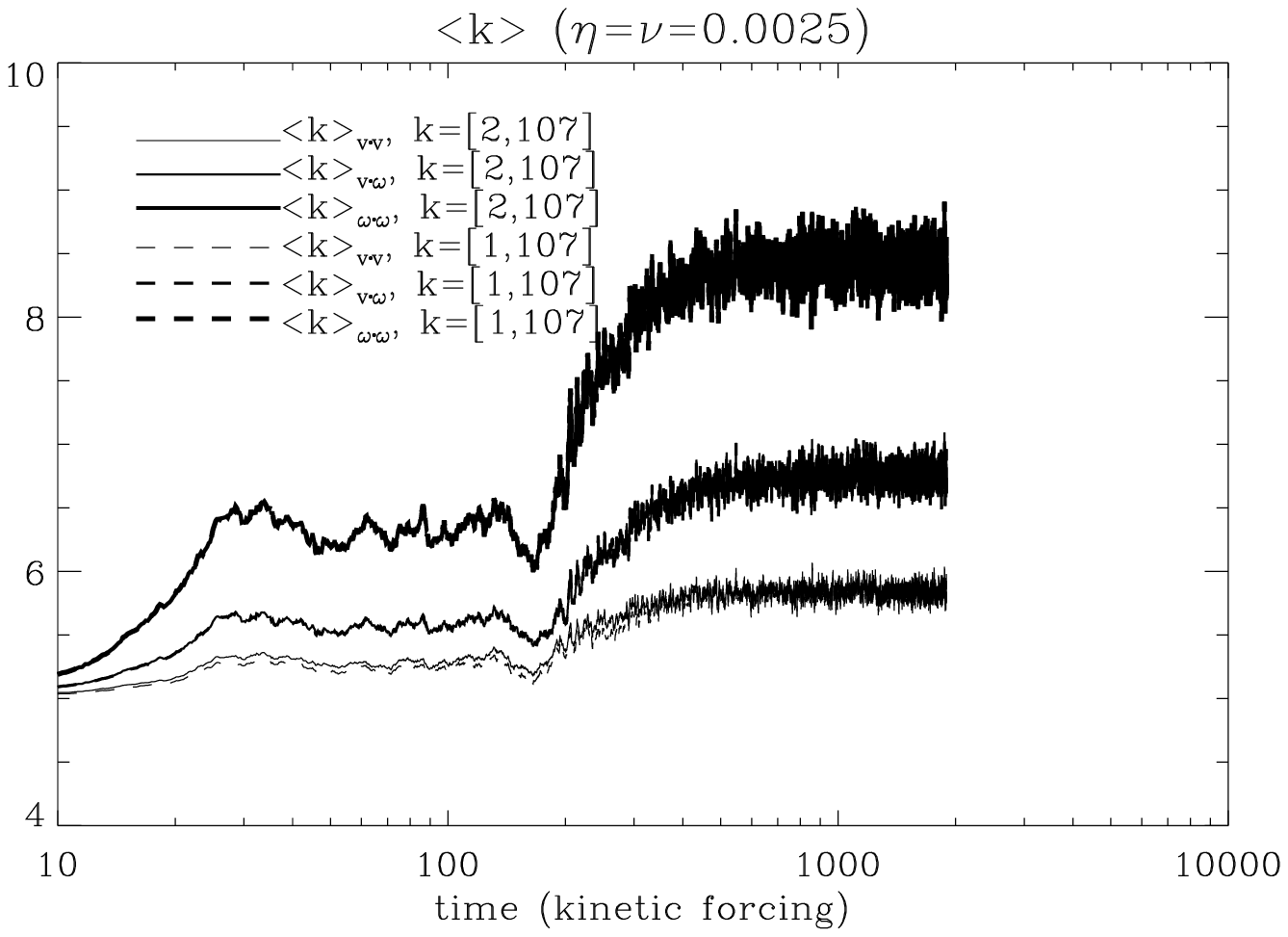}
     \label{kvvvwww0025kf}
   }
}
\caption{Mean kinetic wavenumbers in the different kinetic bases for the two different
resistivity cases for  KF case .  Simulations used a forcing parameter $f_0$=0.07}.
\mbox{%
   \subfigure[]{
     \includegraphics[width=7cm]{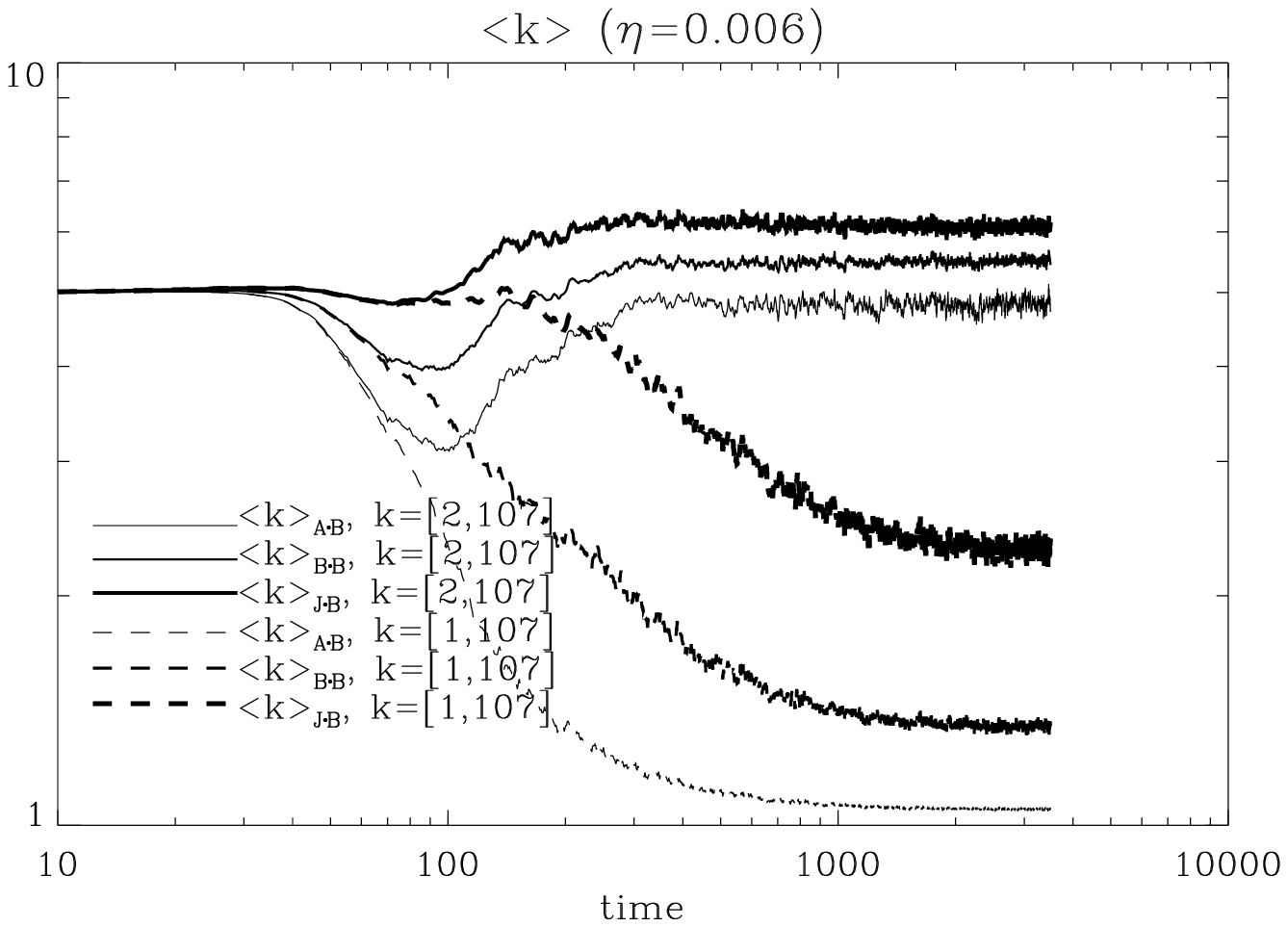}
     \label{kabbbjb006}
   }\,
      \subfigure[]{
     \includegraphics[width=7cm]{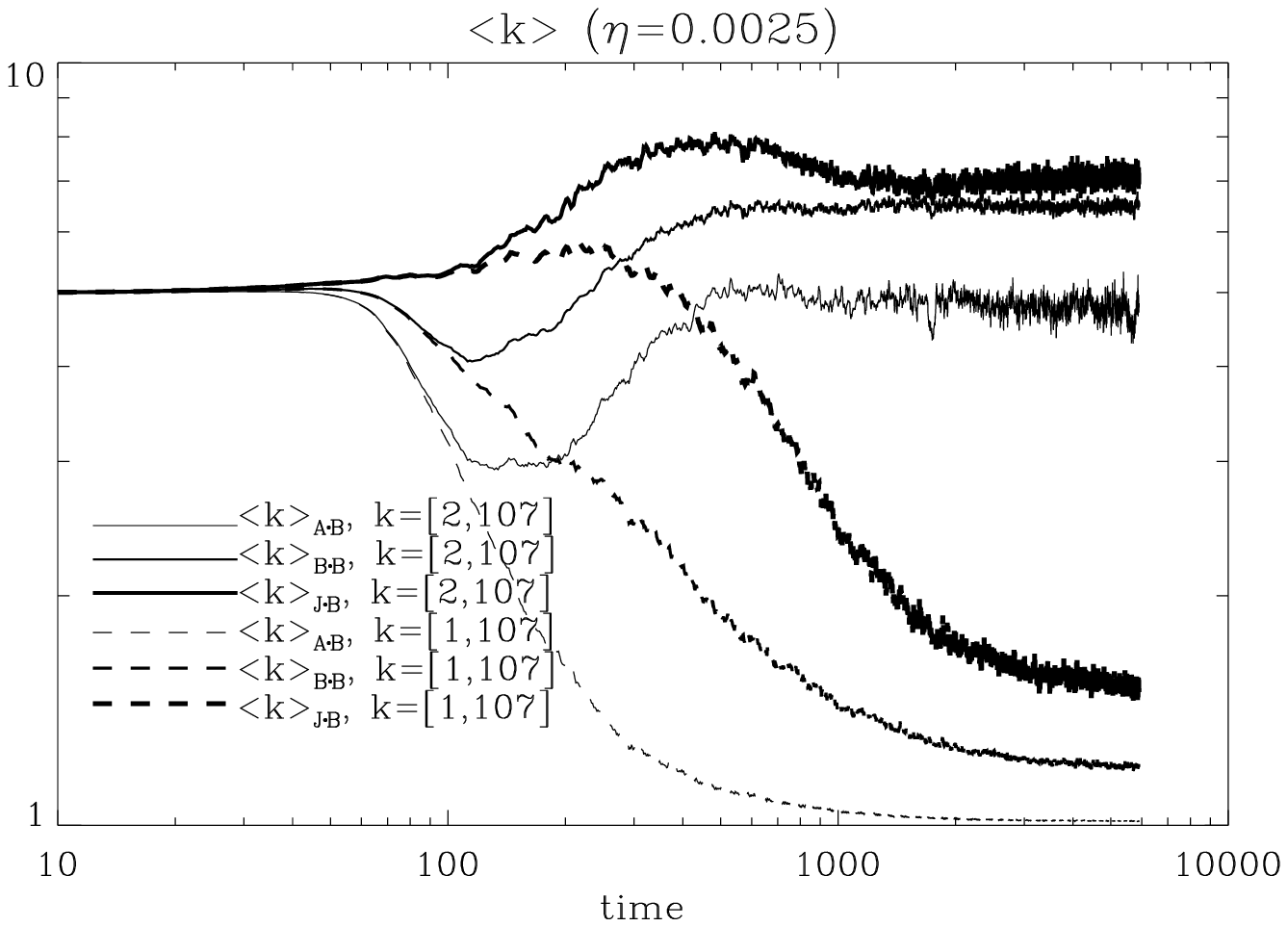}
     \label{kabbbjb0025}
                  }
   }
\caption{Mean magnetic wavenumbers in the different magnetic bases for the two different  resistivity cases for  MF case.  Simulations used a forcing parameter $f_0$=0.03}.
\mbox{%
   \subfigure[]{
     \includegraphics[width=7cm]{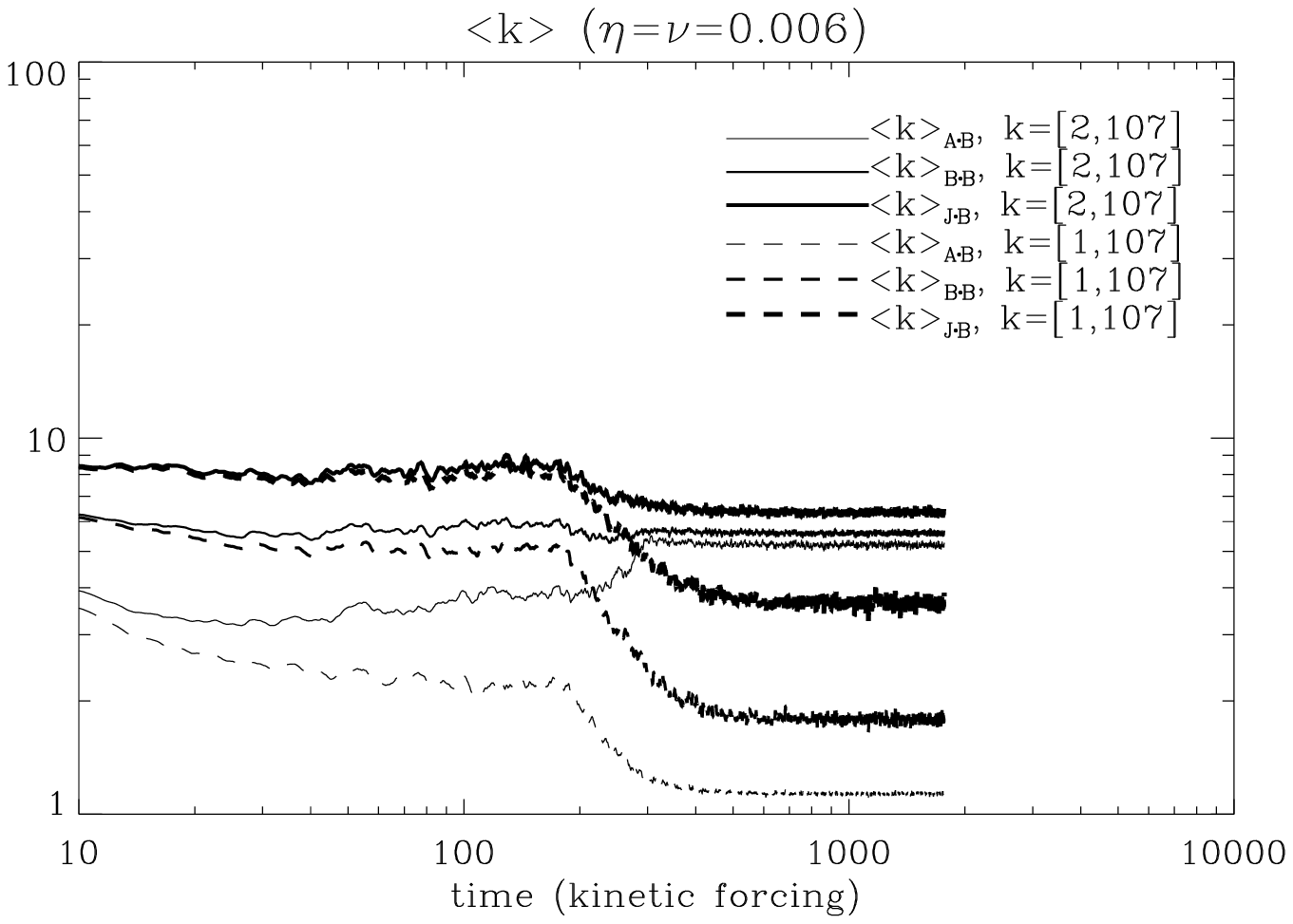}
     \label{kabbbjb006kf}
   }\,
      \subfigure[]{
     \includegraphics[width=7cm]{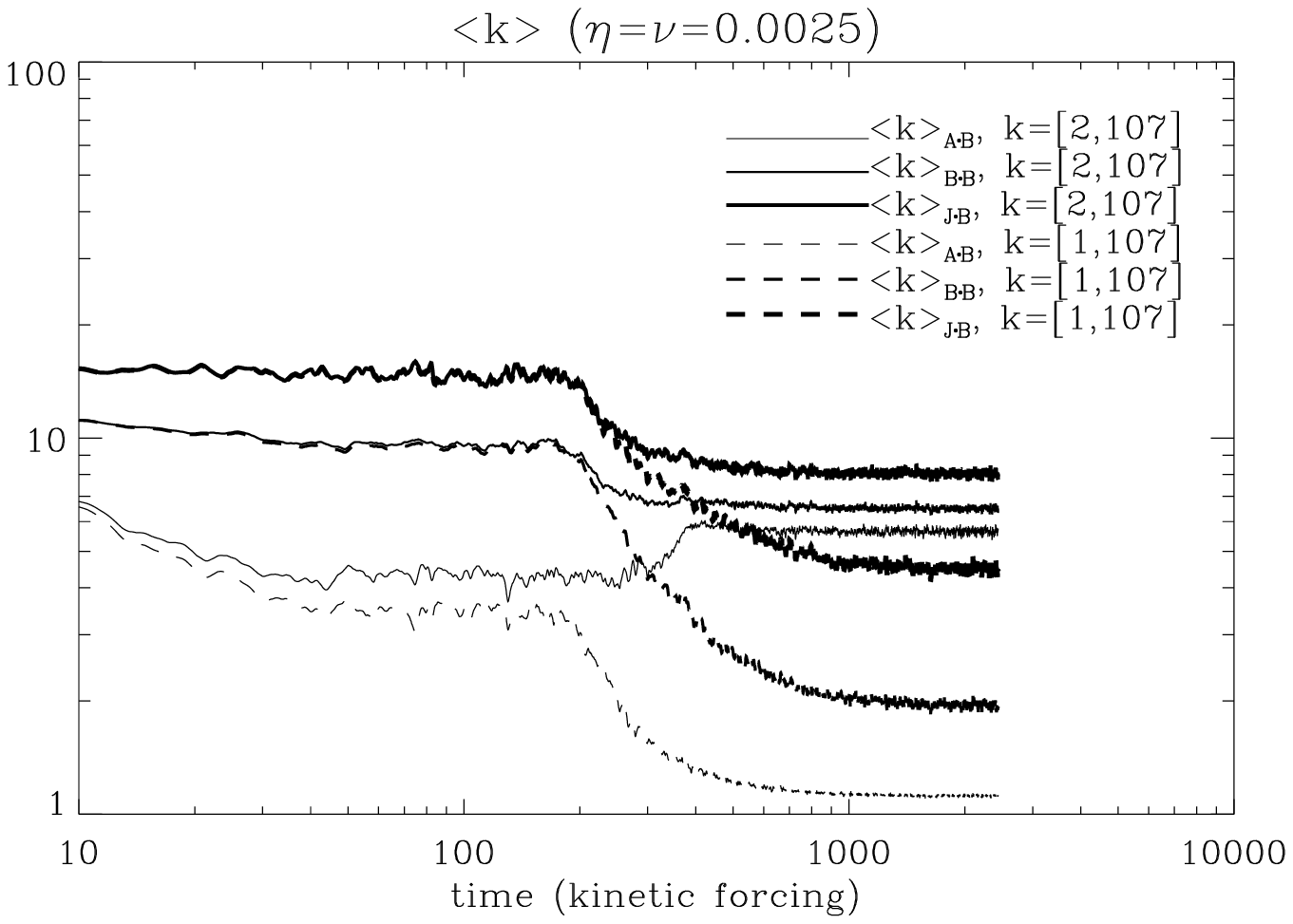}
     \label{kabbbjb0025kf}
                  }
   }
\caption{Mean magnetic wavenumbers in the different magnetic bases for the two different
resistivity cases for  KF case.  Simulations used a forcing parameter $f_0$=0.07}.
\end{figure*}

\begin{figure*}
\centering
\mbox{%
   \subfigure[]{
     \includegraphics[width=8cm]{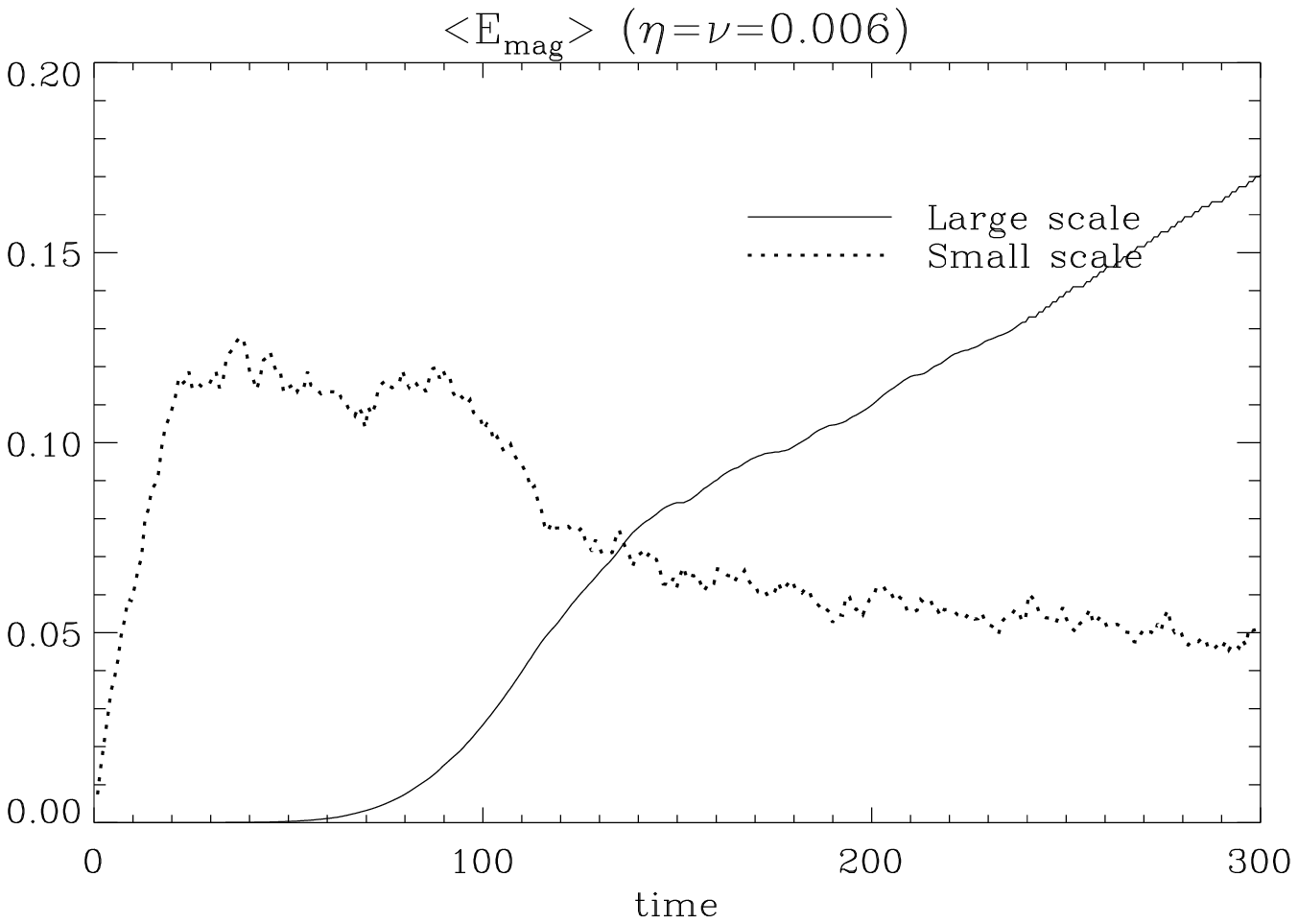}
     \label{emag006early}
   }
   \subfigure[]{
     \includegraphics[width=8cm]{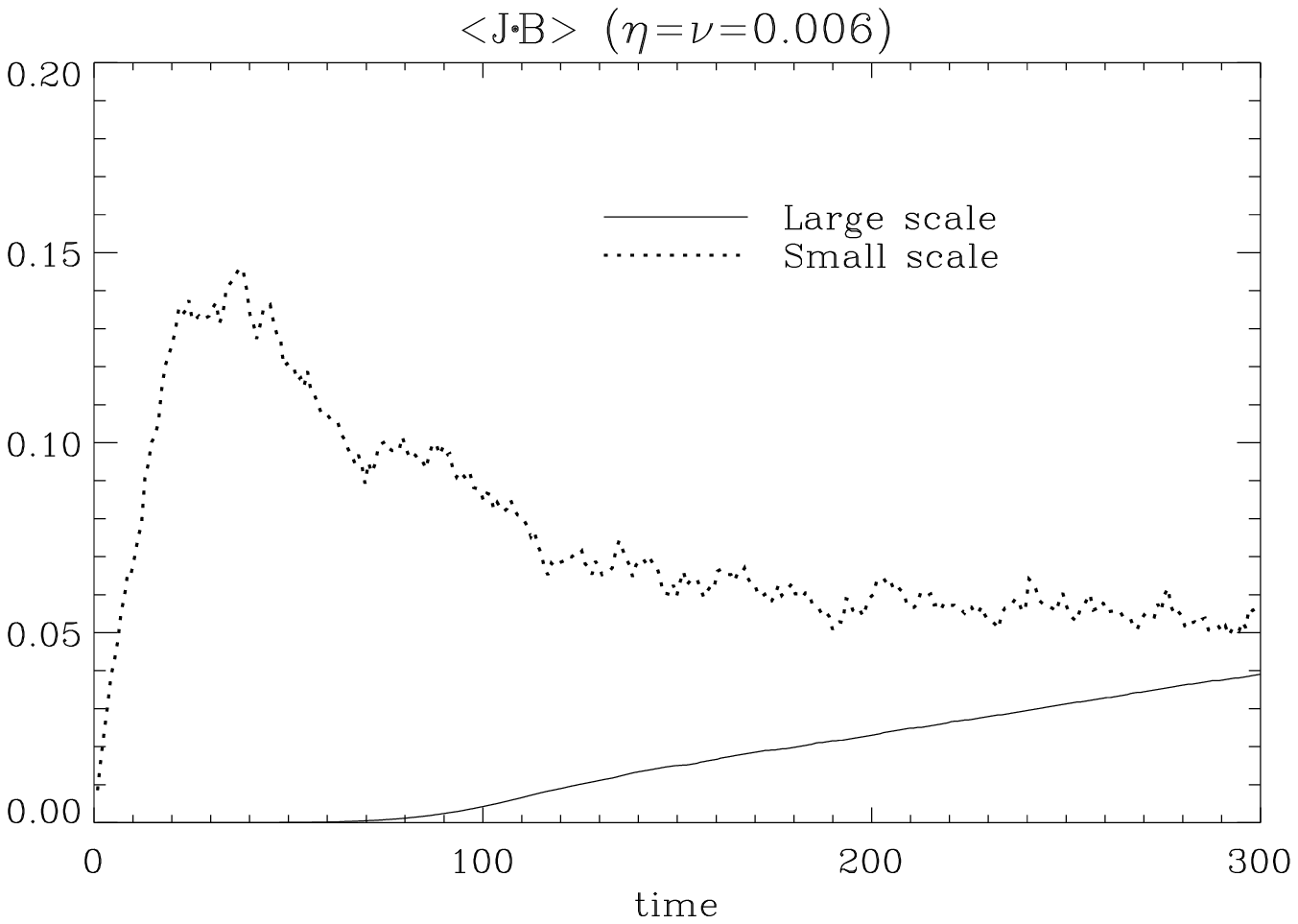}
     \label{jb006early}
   }
}
\mbox{%
   \subfigure[]{
     \includegraphics[width=8cm]{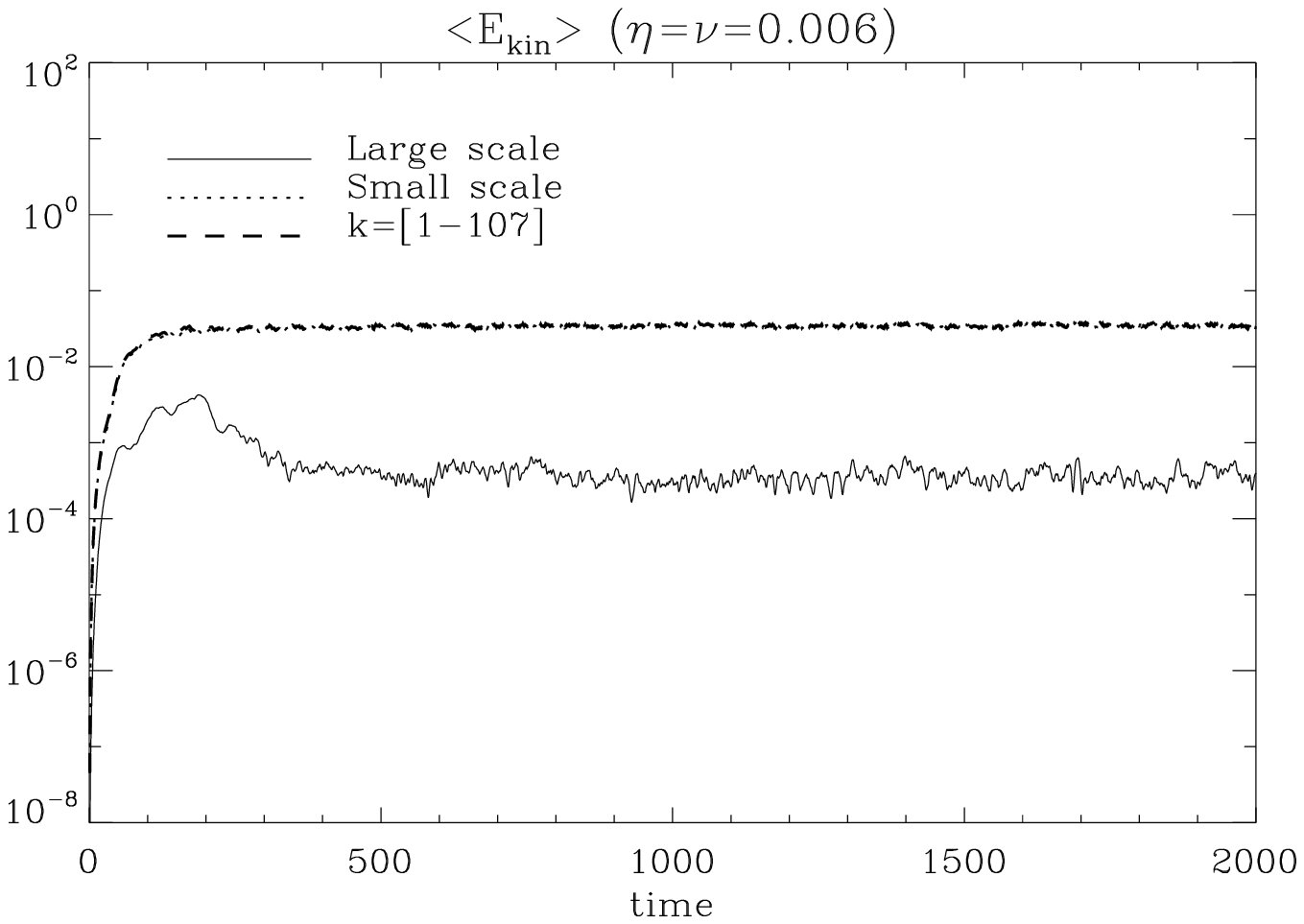}
     \label{eklargesmall006}
   }
   \subfigure[]{
     \includegraphics[width=8cm]{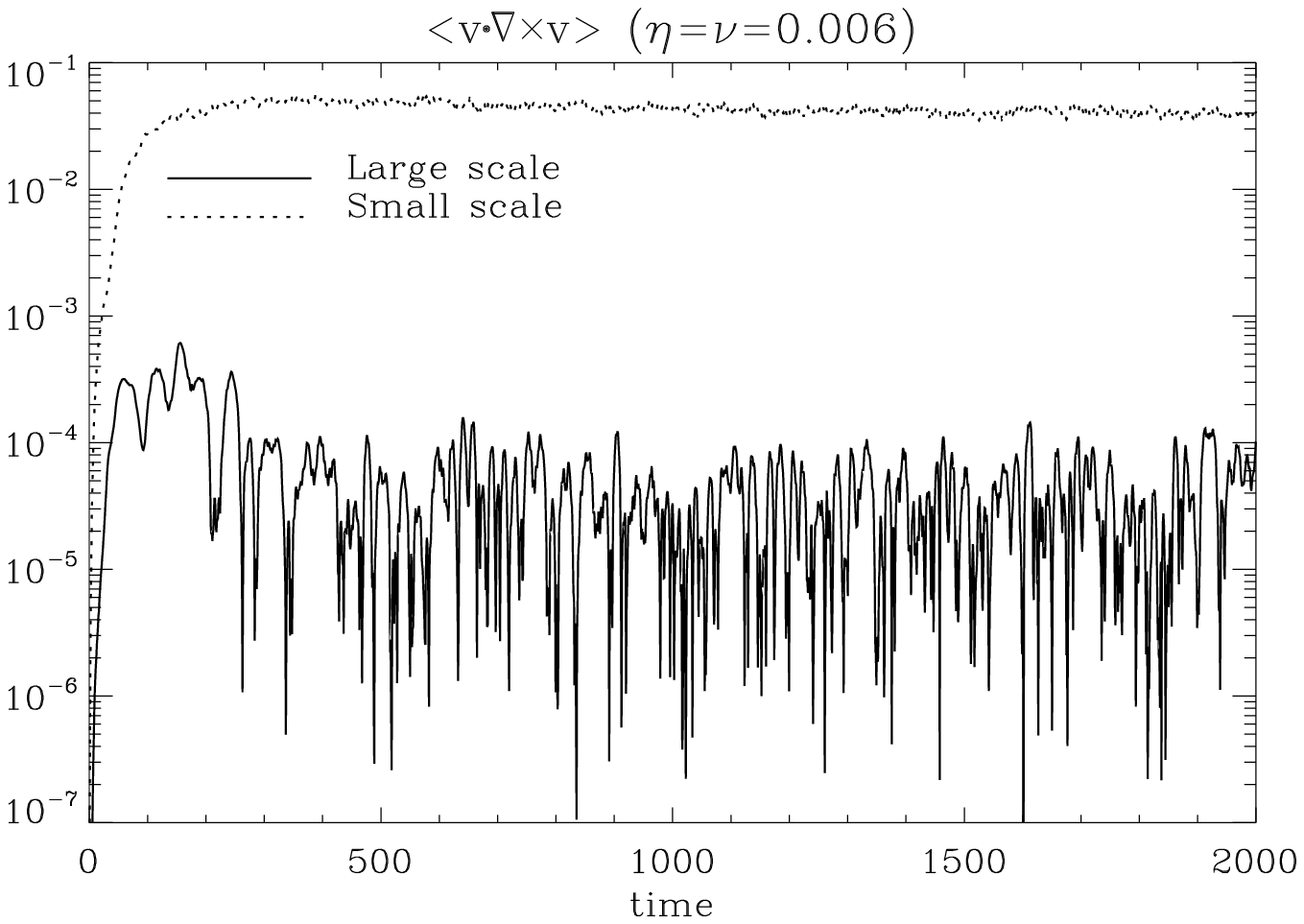}
     \label{vwlargesmall006}}
}
\caption{Time evolution of dimensionless  (a) magnetic energy $E_{mag}$,  (b) current helicity($k_i^2h_i$); (c) kinetic energy(${\bf \varepsilon}$), and (d) kinetic helicity($\langle {\bf v}\cdot \nabla \times {\bf v}\rangle$) in the large and small scale.
The large scale $(k=1)$ kinetic quantities ${\bf \varepsilon}$ and $\langle {\bf v}\cdot \nabla \times {\bf v}\rangle$ are less than 1\% of those in the small scale($k=[2\sim107]$).}
\end{figure*}

%

The equations in this paper were modified from the theory in \cite{2004PhPl...11.3264B}
 with respect to the external force term. In that paper, the forcing was imposed such that
 the small scale magnetic helicity was kept constant from $t=0$.  In the present case, to better match what was done in our simulations, we impose the force in the induction equation.
The modified magnetic induction equation becomes
\begin{eqnarray}\label{induction equation}
\frac{\partial {\bf B}}{\partial t}=\nabla \times \big({\bf V}\times {\bf B}-\eta {\bf J}+{\bf F}\big),
\end{eqnarray}
where  ${\bf F}$ is the forcing function.
We divide  the system into large scale (wave number $k_1$, or $k_L$) and small scale(wave number $k_2$, or $k_s$).

We use an overbar to indicate the mean and lower case letters to indicate fluctuations.
We assume that the forcing applies only to the small scale so that
${\bf F}={\bf f}$ (i.e. ${\overline {\bf F}}=0$),   and is helical such that
$|{\nabla \times \bf f}|= |k_2 {\bf f}|$
We also assume the mean velocity  is zero such that $\bf {V}= \overline{{\bf V}}+v=v$.
We can then write the remaining MHD quantities as
 $\bf {E}= \overline{{\bf E}}+e$, $\bf {B}= \overline{{\bf B}}+b$, $\bf {J}= \overline{{\bf J}}+j$. The  electric field($\bf E=-{\bf V}\times {\bf B}+\eta {\bf J}-{\bf F}$) in large and small scales are
\beq
\label{Efield}
\overline{{\bf E}}=-\overline{\bold{\mathcal{E}}}+\eta\,\bar{{\bf J}},
\eeq
and
\beq
\label{efield}
{\bf e}=-{\bf v}\times\overline{{\bf B}}+\overline{\bold{\mathcal{E}}}+\eta\,{\bf j}-{\bf f},
\eeq
where
$\bold{\mathcal{E}}\equiv\overline{{\bf v}\times{\bf b}}$.

The magnetic helicities in the large and small scale are
\beq
\label{Magnetichelicity1}
\frac{\partial H^M_1}{\partial t}=-2 \langle \overline{{\bf E}} \cdot \overline{{\bf B}}\rangle
=2 \langle \overline{\bold{\mathcal{E}}}\cdot \overline{{\bf B}}\rangle - 2\eta \langle {\bar{{\bf J}}\cdot \bar{{\bf B}}}\rangle
\eeq
and
\beq
\label{Magnetichelicity2}
\frac{\partial H^M_2}{\partial t}=-2 \langle {\bf e} \cdot {\bf b} \rangle\sim -2 \langle \overline{\bold{\mathcal{E}}}\cdot \overline{{\bf B}} \rangle + 2\eta \langle {{\bf j}\cdot {\bf b}}\rangle + 2 \langle {\bf b}\cdot {\bf f}\rangle.
\eeq

\begin{figure*}
\centering
\mbox{%
   \subfigure[]{
     \includegraphics[width=8cm]{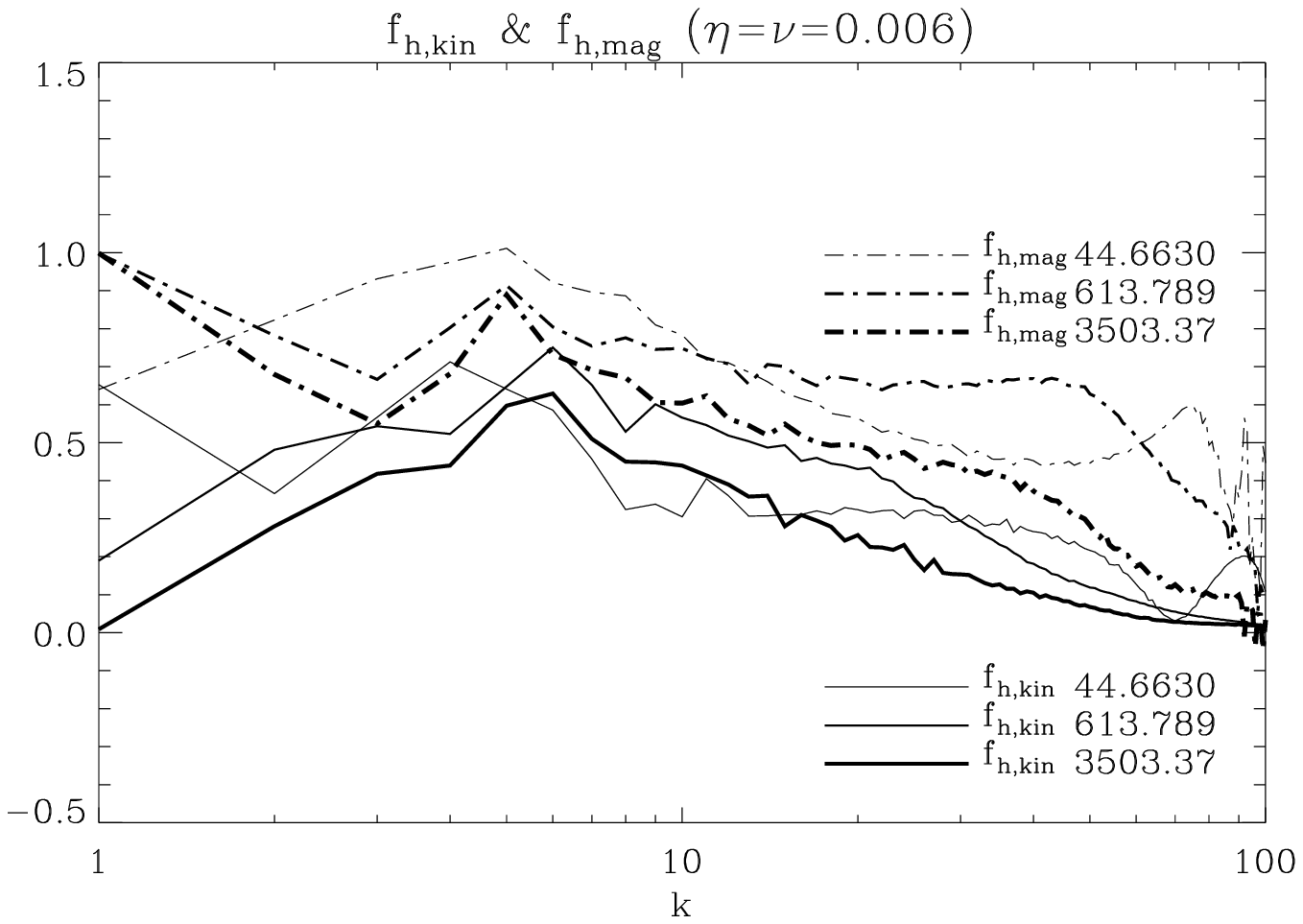}
     \label{fhmfhk006}
   }\,
   \subfigure[]{
     \includegraphics[width=8cm]{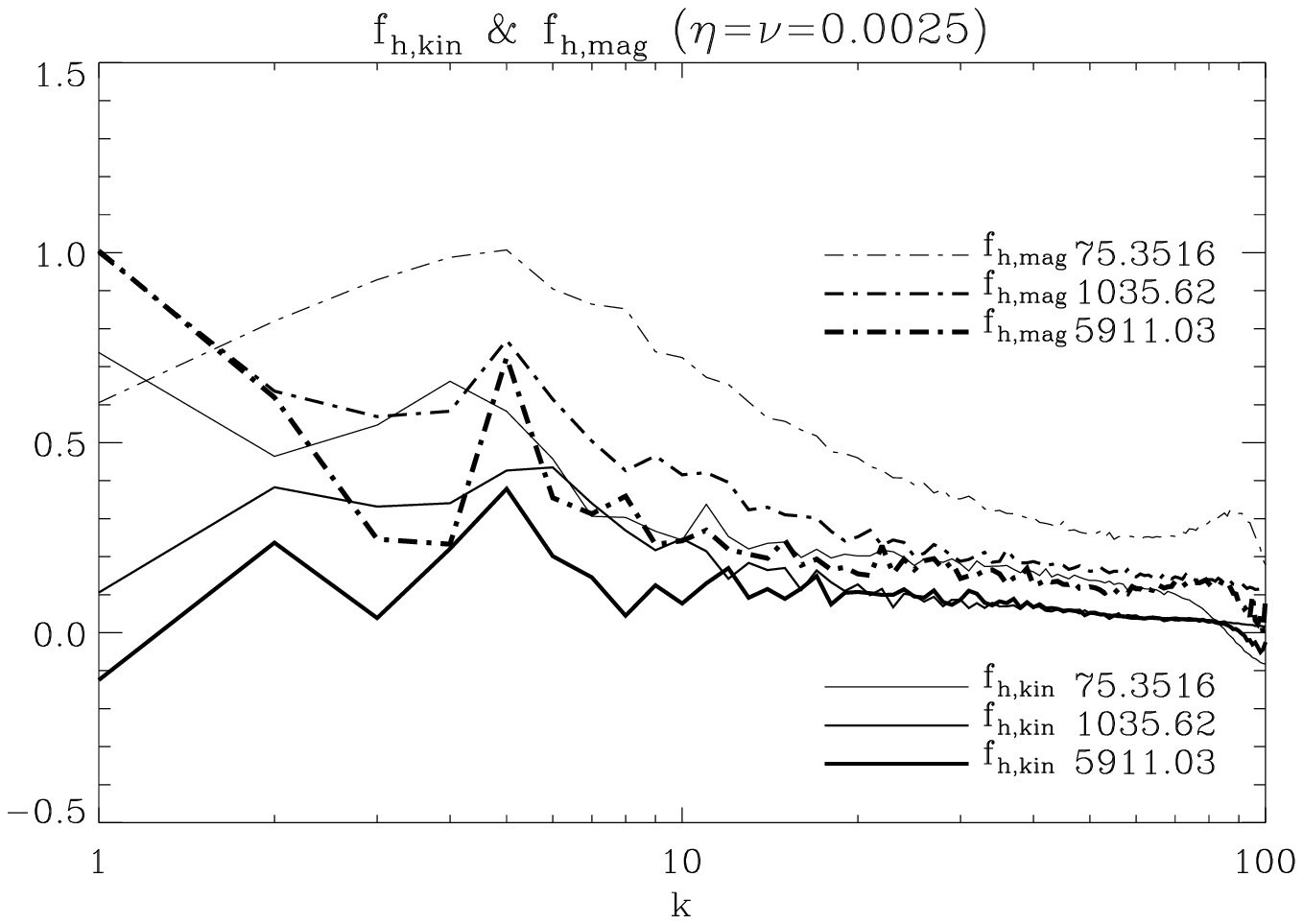}
     \label{fhmfhk0025}
   }
   }
\caption{Comparison of the time evolution of fractional kinetic and fractional magnetic helicity as a function of $k$ for the two different $R_V=R_M$ cases. This
 kinetic energy on the large scale saturates at very low fractional helicities compared to the magnetic case. Note also from the previous figure that there is negligible kinetic energy  on the large scales compared to  magnetic energy; only the fractional helicities are plotted in the present figures.  }
\mbox{%
   \subfigure[]{
     \includegraphics[width=8cm]{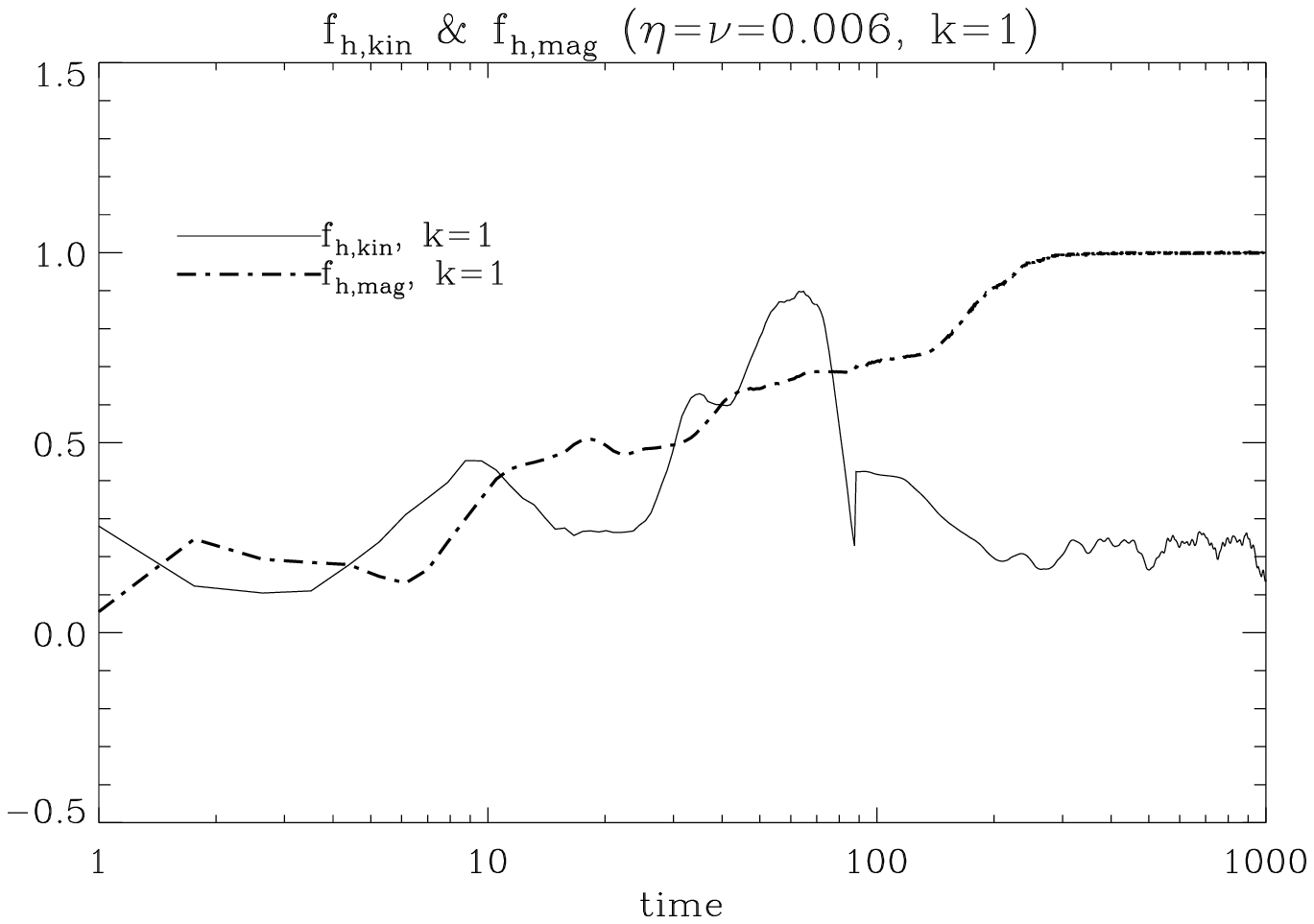}
     \label{fhmfhklargescale006}
   }\,
   \subfigure[]{
     \includegraphics[width=8cm]{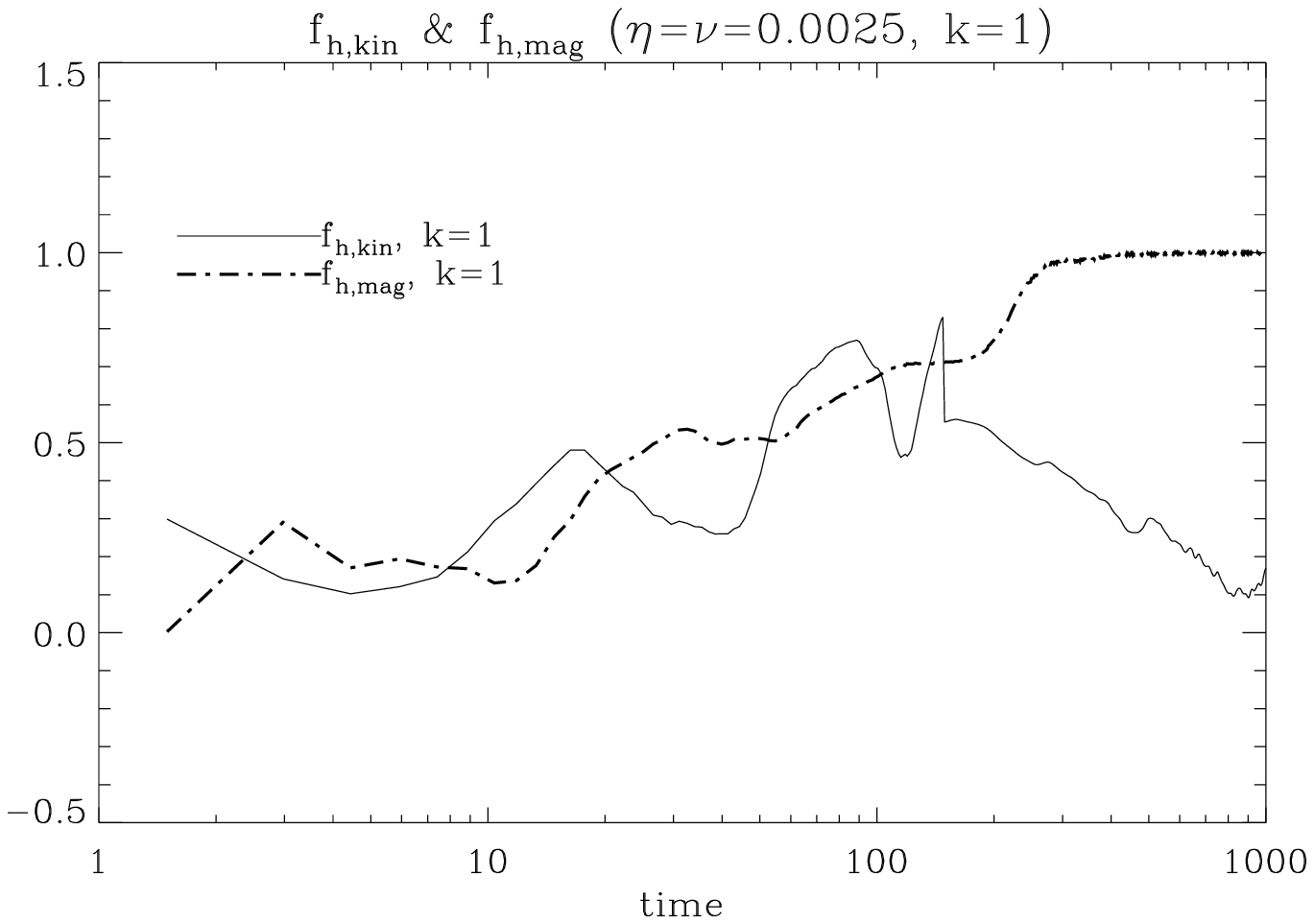}
     \label{fhmfhklargescale0025}
   }
}
\caption{Time evolution of the  fractional helicities $f_{h, kin}$ and $f_{h, mag}$ on the large scale $k=1$.}
\end{figure*}

The equation for $\bold{\mathcal{E}}_{\parallel}$(=$\overline{\bold{\mathcal{E}}}\cdot \overline{{\bf B}} / |\overline{{\bf B|}} $) can be obtained using
\begin{eqnarray}\label{EMF1}
\frac{\partial \bold{\mathcal{E}}_{\parallel}}{\partial t}=\bigg(\overline{\frac{\partial {\bf v}}{\partial t}\times{\bf b}}
+\overline{{\bf v}\times\frac{\partial {\bf b}}{\partial t}}\bigg)\cdot \frac{\overline{\bf B}}{|\bf \overline{B}|}+
\overline{{\bf v}\times{\bf b}}\cdot \frac{\partial}{\partial t}\bigg(\frac{\overline{\bf B}}{|\bf \overline{B}|}\bigg).
\end{eqnarray}
which is derived analogously to that in
 (\cite{1999ApJ...521..597B}, \cite{2002PhRvL..89z5007B}) from  the small scale momentum and magnetic induction equations given by
\begin{eqnarray}
\label{Navier Stokes equation}
\frac{\partial {\bf v}}{\partial t}&=&{\bf v}\times(\nabla\times{\bf v})-\overline{{\bf v}\times(\nabla\times{\bf v})}
-\nabla P_{eff}+{\bf j}\times\overline{{\bf B}}\nonumber\\
&&+\overline{{\bf J}}\times{\bf B}+\nu\nabla^2{\bf v},
\end{eqnarray}
and
\begin{eqnarray}\label{Induction equation3}
\frac{\partial {\bf b}}{\partial t}&=&\overline{{\bf B}}\cdot\nabla{\bf v}-{\bf v}\cdot\nabla \overline{{\bf B}}+{\bf b}\cdot \nabla {\bf v}-\nabla \times \overline{\bold{\mathcal{E}}}+\lambda \nabla^2{\bf b}+ k_2 \, {\bf f}.\nonumber\\
\end{eqnarray}
The equation for the EMF $\bold{\mathcal{E}}$ becomes,
\begin{eqnarray}\label{EMF2}
\frac{\partial \bold{\mathcal{E}}_{\parallel}}{\partial t}&=&\frac{1}{3}\bigg(\overline{{\bf b}\cdot \nabla \times {\bf b}}-\overline{{\bf v}\cdot \nabla \times {\bf v}}\bigg)\frac{{\bf \overline{B}}^2}{|{\bf \overline{B}}|}
-\frac{1}{3}\overline{{\bf v}}^2\nabla \times \overline{{\bf B}} \cdot \frac{{\bf \overline{B}}}{|{\bf \overline{B}}|}\nonumber\\
&&-c\, k_2\, \overline{{\bf v}^2}^{1/2}\bold{\mathcal{E}}_{\parallel} + k_2 \, \overline{{\bf v}\times {\bf f}}\cdot \frac{{\bf \overline{B}}}{|{\bf \overline{B}}|}.
\end{eqnarray}
Here the term involving  $c\, k_2\, \overline{v^2}^{1/2}\bold{\mathcal{E}}_{\parallel}$ is the result of  the ``minimal tau'' closure, used e.g \cite{2002PhRvL..89z5007B} that incorporates
 triple correlations `$T$' as well as microscopic magnetic diffusivity `$\eta$', kinetic viscosity `$\nu$', triple and the last term (Eq.(6) in \cite{2002PhRvL..89z5007B}). The quantity $c$ is a constant that will later be determined empirically.\\
\\
The mean kinetic energy per unit mass ($\overline{v^2}$/2$=\langle v^2\rangle$/2) can be obtained from the momentum equation,
\begin{eqnarray}\label{U1}
\frac{\partial \langle v^2\rangle}{\partial t}&\sim& 2\langle {\bf v}\cdot {\bf j} \times \overline{{\bf B}}\rangle+2\langle {\bf v}\cdot \overline{{\bf J}} \times {\bf b}\rangle+2\nu \langle {\bf v}\cdot \nabla^2{\bf v}\rangle\nonumber\\
&=& 2\langle ({\bf v}\times {\bf j}) \cdot \overline{{\bf B}}\rangle - 2\langle ({\bf v}\times {\bf b})\cdot \overline{{\bf J}} \rangle - 2\nu \langle (\nabla {\bf v})^2\rangle.
\end{eqnarray}
Using vector identities and defining the helicity ratios (\cite{2012MNRAS.419..913P}),
\begin{eqnarray}\label{Magnetic Helicity Ratio}
f_{mi}&=&\bigg\langle \frac{k_i{\bf A}_i\cdot {\bf B}_i}{|{\bf B}_i|^2}\bigg\rangle=\bigg\langle\frac{{\bf J}_i\cdot {\bf B}_i}{k_i|{\bf B}_i|^2}\bigg\rangle
\end{eqnarray}
\begin{eqnarray}\label{Kinetic Helicity Ratio}
f_{hi}&=&\bigg\langle\frac{{\bf v}_i\cdot\nabla \times {\bf v}_i}{k_i v_i^2}\bigg\rangle\quad (i=1, 2),
\end{eqnarray}
we  find the equation for $\langle v^2 \rangle$ to be
\begin{eqnarray}\label{U2}
\frac{\partial \langle v^2 \rangle}{\partial t}\sim 2(f_{m2}k_2-f_{m1}k_1)\langle \overline\bold{\mathcal{E}}\cdot \overline{\bf B}\rangle -2\nu k_2^2 \langle v^2 \rangle.
\end{eqnarray}

 We nondimensionlize  the equation with the scalings  below:
\begin{eqnarray}\label{dimensionless coefficients}
&&H_1^M \equiv h_1 H_{2s}, \, H_2^M \equiv h_2 H_{2s}, \, \tau \equiv t k_2^{3/2}\sqrt{H_{2s}}, \nonumber\\
&&Q \equiv \overline{\bold{\mathcal{E}}}_{\parallel}/k_2 H_{2s}, \varepsilon \equiv U/k_2 H_{2s}, R_M \equiv H^{1/2}_{2s}/\eta k_2^{1/2}, \nonumber\\
&&R_V \equiv H^{1/2}_{2s}/\nu k_2^{1/2}, H_{2s}\equiv b_{rms}^2/k_2.
\end{eqnarray}
The use  of $k_2$ and root mean squared small scale magnetic field $b_{rms}(\equiv b_r)$ in these normalizations
is natural since the external MF was applied to  $k=k_2=5$, and the $RMS$
small scale field is steadily injected.
The  ordinary differential equations to be simultaneously solved are the dimensionless versions of
Eq.(\ref{Magnetichelicity1}), (\ref{Magnetichelicity2}), (\ref{EMF2}), and (\ref{U1}), which are given respectively by
\begin{eqnarray}\label{dimensionless equations1}
\frac{\partial h_1}{\partial \tau}=\overbrace{\frac{2}{f_{m1}^{1/2}}\big(\frac{k_1}{k_2}\big)^{1/2}Q h_1^{1/2}}^{1 \,(\bold{\mathcal{E}}_{s\rightarrow L})}\overbrace{-\frac{2\eta}{b_r}\frac{k_1^2}{k_2}h_1}^2,
\end{eqnarray}
\begin{eqnarray}\label{dimensionless equations2}
\frac{\partial h_2}{\partial \tau}&=&\overbrace{-\frac{2}{f_{m1}^{1/2}}\big(\frac{k_1}{k_2}\big)^{1/2}Q h_1^{1/2}}^{3\, ( -\bold{\mathcal{E}}_{s\rightarrow L})}
\overbrace{-\frac{2\eta}{b_r}k_2 h_2}^{4}+\overbrace{\frac{2}{b_r^2}\frac{1}{f_{m2}^{1/2}}h_2^{1/2}f'
}^{5\,(forcing)},\nonumber\\ 
\end{eqnarray}
\begin{eqnarray}\label{dimensionless equations3}
\frac{\partial Q}{\partial \tau}&=&\overbrace{\frac{1}{3f_{m1}^{1/2}}\big(h_2-f_{h2}\epsilon \big)\big(\frac{k_1}{k_2}\big)^{1/2}h_1^{1/2}}^{6\,(\alpha\, \sim \tau(\langle \bf b\cdot \nabla \times \bf b\rangle -\langle \bf v \cdot \nabla \times \bf v \rangle)) }\overbrace{-\frac{1}{3}f_{m1}^{1/2}\big(\frac{k_1}{k_2}\big)^{3/2}\,
\epsilon \, h_1^{1/2}}^{7\, (\beta \sim \langle v^2\rangle)}\nonumber\\
&&\overbrace{-c\, \epsilon^{1/2}\,Q}^{8\, (micro\, diss.)}+\overbrace{\frac{\epsilon^{1/2}}{b_r^2}f'}^{9}(\sim0)
\end{eqnarray}
and
\begin{eqnarray}\label{dimensionless equations4}
\frac{\partial \epsilon}{\partial \tau}&=&\overbrace{\frac{2}{f_{m1}^{1/2}}\big(f_{m2}-f_{m1}\frac{k_1}{k_2}\big)\big(\frac{k_1}{k_2}\big)^{1/2}
h_1^{1/2}Q}^{10\,(\langle (\bf v \times \bf j)\cdot \bar{\bf B}\rangle -\langle (v\times b)\cdot \bar{\bf J}\rangle)}\overbrace{-\frac{2\,\nu}{b_r}k_2\,\epsilon}^{11}.
\end{eqnarray}
\\
In our analytic two-scale model we have
assumed that the  kinetic helicity of the fluctuations
satisfies  $\langle \bfv \cdot\nabla\times\bfv \rangle = f_{h2} k_2 \langle v^2 \rangle$
with $f_{h2}$ fixed and $k_2 =k_f=5$. Here the sign of the small scale kinetic helicity is positive, which  is consistent with our simulations expected when $H_2^M$ is initially positive from the imposed forcing.
Nondimensionally, we have   $h_2= f_{h2} \epsilon$.
 Although the simulations show that $f_{h2}$ at $k=5$ is not
strictly constant at all  times during the simulations, the assumption of such is not too far off.
In more accurate treatments a separate dynamical equation for $h_2$ should be included.

Note also that the term labeled 5 on the right of Eq.(\ref{dimensionless equations2})
resulted from  $\langle {\bf b} \cdot {\bf f} \rangle$
and was not solved for dynamically. We empirically determined
the magnitude of $f'$, the magnitude of the forcing needed in the theoretical equations to best match the simulations which employed a forcing magnitude $f_0$.  The magnitude of forcing used in simulation $f_0=0.03$  is slightly larger than the
values of  $f'$ employed in the two-scale model (see Table 1).

The term  labeled 9  in Eq.(\ref{dimensionless equations3})
namely $\langle \overline{{\bf v} \times {\bf f}}\rangle\cdot {\bf \overline{B}\over \overline{B}}$ was taken to be zero: it requires a correlation between two nearly isotropic functions. This assumption
is consistent with the simulations as the quantity is measured to be even smaller than $\langle\bfv\times \bfb\rangle$, which also requires a correlation between two nearly isotropic functions.

\subsection{Discussion of Solutions}

We numerically solve the ordinary set of differential equations Eq.(\ref{dimensionless equations1}), (\ref{dimensionless equations2}), (\ref{dimensionless equations3}), and (\ref{dimensionless equations4}) for
$h_1$, $h_2$, $Q$ and $\epsilon$ (from which we obtain $h_v$).
Table 1 shows the parameters  defined in the previous section that best match the simulation. The dotted lines   in Figs
\ref{h1h2006}, \ref{h1h20025}, \ref{h1006and0025}, \ref{eqhv006}, \ref{eqhv0025}
represent the solutions with these  parameter choices.

\begin{table}
\begin{tabular}{|c|c|c|c|c|c|c|}
\hline
$\eta=\nu$ & $v_{r}$ & $Re_M$ & $f_{m1}$ & $f_{m2}$ & $f_{h2}$ & $b_r$  \\
\hline
0.006 & 0.23 & 48 & 1 & 0.615 & 0.645 & 0.88   \\
\hline
0.0025 & 0.28 & 141 & 1 & 0.4 & 0.185 & 1.5   \\
\hline

\hline
 & $c$ & $f'$ & $Q_0$ & $h_{10}(10^{-14})$ & $h_{20}$ & $\epsilon_0(10^{-8})$\\
\hline
0.006 & 0.45 & 0.02 & 0.042 & 1.23 & 0.015 & 9.3\\
\hline
0.0025 & 0.8 & 0.0195 & 0.022 & 0.062 & 0.0054 & 5.7\\
\hline
\end{tabular}
\label{valuesofvariable}\\
\caption{Table of parameter values used in the analytic equations to best match
the simulations.  Initial values $Q_0$, $h_{10}$, and $h_{20}$
are dimensionless values(Eq.(\ref{dimensionless coefficients})). We use $k_1=1$ and $k_2=5$.}
\end{table}

From Eq.(\ref{dimensionless equations3}) for $Q$, we see that the $\alpha$ coefficient (term 6) increases at first  due to  $h_2$ supplied by the forcing. This drives the growth of $h_1$ of the same sign and  the kinetic energy $\epsilon$ also grows.
Some of this kinetic energy is helical and so $h_v$ grows. This
$h_v$ grows with the  same sign of $h_2$ so when the former is subtracted from the latter   in $\alpha$
the overall alpha effect is reduced.   Thus $h_v$  acts as the  back reaction in $\alpha$.

The associated buildup of $\epsilon$ also increases the turbulent diffusion of $Q$
(7) and the dissipation term (8).
Note that the growth of the turbulent diffusion coefficient as
a backreaction in the MF case differs from that of the KF where the driving itself directly supplies steady turbulent diffusion coefficient from the outset.
The turbulent diffusion has no magnetic component and so must wait for the velocity field to build up.
This delay in growth of the turbulent diffusion is consistent with
the difference between the early time evolution of the mean magnetic basis wavenumbers  in
the MF case  for Fig.\ref{kabbbjb006} and \ref{kabbbjb0025} compared to
KF case  of Fig.\ref{kabbbjb006kf} and  \ref{kabbbjb0025kf}. In the latter,
the KF immediately leads to a turbulent cascade available to induce scale separation
in the magnetic quantities.

As the backreaction from the kinetic energy and kinetic helcity grow,
$dQ/dt$ converges to zero and $Q$ becomes saturated at small positive value after the initial bump(Fig.\ref{eqhv006}, \ref{eqhv0025})
The right hand side(RHS) of Eq.(\ref{dimensionless equations1})(terms 1+2) then also converges to zero and $h_1$ saturates. The behavior of $h_2$ can be explained  similarly.

Complete saturation occurs when all of the terms on the right sides of the dimensionless equations Eq.(\ref{dimensionless equations1}), (\ref{dimensionless equations2}),
(\ref{dimensionless equations3}), and (\ref{dimensionless equations4}) equal zero. However, the largest backreaction term
is the kinetic helicity term and so setting $h_2 \sim h_v$ gives an approximate
assessment of  quenching.  This approximate equality is consistent with
the asymptotic values of $h_2$ and $h_v$ shown in Fig.\ref{h1h20025} and \ref{eqhv0025}, given that
$k_2=5$. The late time growth is limited by viscosity in the sense that the faster $h_v$ viscously dissipates, the less effective its backreaction is on $h_2$. This is evident in Fig.\ref{h1006and0025} which shows that the lower $R_V=R_M$ case approaches saturation earlier than the $R_V=R_M$ case.

It is important to emphasize that it is the viscosity
and not the resistivity that limits the MF LSD. This contrasts the KF
case which is resistively limited by how fast  the small scale magnetic helicity resistively dissipates relative to the large scale magnetic helicity.  In that respect, keeping  $R_M$ fixed while reducing $R_V$ increases the magnetic Prandtl number and the saturation rate of the MF case. This was discussed in \cite{2004PhPl...11.3264B}. We do not explore the Prandtl
number dependence in the present paper.  In general, future work will be needed to better understand the early time regime and the associated $R_M$ and $R_V$ dependence.

.

\subsection{Analytic Simplifications}

Although Eq.(\ref{dimensionless equations1}), (\ref{dimensionless equations2})
(\ref{dimensionless equations3}), and (\ref{dimensionless equations4})
require a numerical solution, some insight can be gained by an analytic approximation. The $h_1$ equation has the form of a   Bernoulli equation
and can be linearized by changing variables. If we use `$u_1$'$(\equiv h_1^{1/2})$, the differential equation for $h_1$ becomes,
\begin{eqnarray}\label{linearized h1}
\frac{\partial u_1}{\partial \tau}&=&-\frac{\eta}{b_r}\frac{k_1^2}{k_2}u_1+\frac{1}{f_{m1}^{1/2}}\big(\frac{k_1}{k_2}\big)^{1/2}Q
\equiv -P_1\, u_1+Q_1.\\
&& \big(P_1\equiv \frac{\eta}{b_r}\frac{k_1^2}{k_2},\quad Q_1\equiv \frac{1}{f_{m1}^{1/2}}\big(\frac{k_1}{k_2}\big)^{1/2}Q,\quad \big)\nonumber
\end{eqnarray}
The solution of this standard inhomogeneous differential equation is  known. Using the same method as in \cite{2012MNRAS.419..913P} we find,
\begin{eqnarray}\label{approximate solution of h1a}
u_1(t)&=&e^{-\int^t P_1 dt'}\bigg[\int^t_0 e^{\int^{t'} P_1 dt''}Q_1(t')dt'+Const \bigg]\nonumber\\
&=&e^{-P_1\,t}\bigg[\int^t_0 (e^{P_1\,t})'\frac{Q_1(t')}{P_1}dt'+Const\bigg]\nonumber\\
&\sim & \frac{Q_1(t)}{P_1}+\bigg(Const-\frac{Q_1(0)}{P_1}\bigg)e^{-P_1\,t}.
\end{eqnarray}
As $t\rightarrow \infty$, `$u_1(t)(\equiv h_1^{1/2})$' approaches to `$Q_1(t)/P_1$'($=\overline{\bold{\mathcal{E}}}(t)k_2^{1/2}/f_{m1}^{1/2}b_r\eta k_1^{3/2}$) which can be  calculated by setting `$dh_1/dt$' to be zero.  This approximate value `$Q_1(t)/P_1$' matches the numerically solved $h_1^{1/2}$ after $t\sim 1500$, for $\eta=\nu=0.006$.\\

The equation for $h_2^{1/2}$ can not be linearized. Instead, setting $dh_1/dt=0$, and $u_2\equiv h_2^{1/2}$ we obtain a quadratic equation:
\begin{eqnarray}\label{approximate solution of h2}
u_2(t)&=&\frac{1}{2}\bigg[\frac{f'}{\eta k_2 b_r f_{m2}^{1/2}}-\sqrt{\bigg(\frac{f'}{\eta k_2 b_r f_{m2}^{1/2}}\bigg)^2-\frac{4b_r Q h_1^{1/2}}{\eta f_{m1}^{1/2}}\big(\frac{k_1}{k_2^3}\big)^{1/2}}\bigg].\nonumber\\
\end{eqnarray}

Except the early time regime($t\lesssim 130$, $\eta=\nu=0.006$), this solution fits the numerical result well. These approximate equations explain how the saturated magnetic helicities in large and small scale depend on `$\eta$'.
\\

\section{Astrophysical Relevance of Magnetically Forced Large Scale Dynamos}

The MF LSD that we have discussed in this paper maintains a
plasma $\beta_p$ greater than unity for most of the time.
Although we consider the simplest such MF LSD in that we employ periodic boundaries and no rotation or shear, the basic concept of a LSD that is driven by small scale magnetic fluctuation rather than kinetic fluctuation in $\beta_p >1$ environments has important conceptual relevance to both solar dynamos and dynamos in accretion discs.

In the solar context, helical LSD models have been  separated into two classes: (1)
 helical forcing is primarily kinetic helicity driven by thermal convection (2)
 flux transport models which  helical fields are the result of  magnetic instabilities (cf. \cite{2007AdSpR..39.1661C}).
We suggest that the second class of dynamos can be viewed as an MF LSD, albeit with more complexity than the simple version we consider in the present paper.
Magnetic instability driven dynamos in the radiative zone (\cite{2002A&A...381..923S})
also seem to fit into this category.

For accretion discs, large scale dynamos are now commonly seen in shearing box simulations(e.g. \cite{1995ApJ...446..741B, 2010ApJ...713...52D,
2010MNRAS.405...41G, 2011MNRAS.413..901K}).  The magneto-rotational instability operating in these simulations produces turbulence that drives an LSD. But the
the magnetic fluctuations typically exceed the kinetic fluctuations and
it seems that the best agreement between mean field theory and simulation
requires that the driver of the LSD growth is not the kinetic helicity term but the current helicity term (\cite{2010MNRAS.405...41G}).  Thus accretion discs are another environment where more general $\beta_p>1$ MF LSD dynamos operate.

Although the calculations of the present paper are in the $\beta_p \gtrsim 1$ limit,
we note that for the $\beta_p \ll 1$ limit, MF LSD have long been studied in the
laboratory context (e.g. \cite{2002MHD....38..191J}). The direct analogy to these low $\beta$ MF LSDs may occur in astrophysical coronae (\cite{2007NJPh....9..309B}).

\section{Conclusion}

We performed numerical simulations of the analogue of an $\alpha^2$ LSD in a periodic box when the system is driven with magnetic (or current) helicity rather than kinetic helicity.  The simulations indeed show that LSD action  results from the injection of small scale magnetic helicity,  analogously to the KF LSD action from injection of kinetic helicity. This analogy is expected because the growth driver is proportional to the time integrate difference between current helicity and kinetic helicity.

We compared the simulation results with a  two scale theory and found general consistency with respect to the basic mechanism of large scale field growth and saturation. When the system is magnetically driven at $k=5$,  the large scale $k=1$ helical magnetic field  grows in the MF with the same sign as that on the driving scale.
Injecting  $k=5$ magnetic helicity drives the
system away from its natural relaxed state and the LSD  evolves the magnetic helicity
to the large scale where the same amount of magnetic helicity has lower energy.
Eventually the LSD saturates because the growth driver is the difference between
kinetic and current helicities and the kinetic helicity builds up to quench the LSD.
This situation complements the KF forced case where the LSD drives oppositely signed growth of large and small scale magnetic helicities and the current helicity quenches the LSD.

We presented  the time evolution of the spectra of magnetic energy, magnetic helicity,  current helicity from the simulations. Taken together, these spectra exhibit the expected inverse cascade of magnetic helicity that is at the heart of LSD action, and the absence of an inverse cascade of kinetic helicity. Future work is needed to better understand the early time growth regime and study the  dependence on Reynolds number and Prandtl numbers.

The MF LSD studied herein was for MHD plasmas with ratios of thermal to magnetic pressure overall larger than unity. We checked that this was the case at all times
during the simulations.  We discussed that such an MF LSD may ultimately be involved in producing the LSD action in MRI simulations and may also be away to distinguish
Babcock type solar dynamo models from KF kinetic helicity driven solar dynamo models.



\section{Acknowledgement}
We acknowledge support from US NSF grants PHY0903797 and AST1109285. KWP acknowledges a Horton Fellowship
from the Laboratory for Laser Energetics at the University of Rochester. We also appreciate the suggestion and help of
Dr. Axel Brandenburg and Dr. Brendan Mort.

\bibliographystyle{mn2e}
\bibliography{bibfile}
\end{document}